\documentclass{aa}  

\usepackage{graphicx}
\usepackage{txfonts}
\usepackage{natbib}
\bibpunct{(}{)}{;}{a}{}{,} 

\begin{document}

\title{Spectral line lists of a nitrogen gas discharge for wavelength calibration in the range $4500-11000$\,cm$^{-1}$}
\author{A. Boesch
       \and
       A. Reiners
       }
\institute{Institut f\"ur Astrophysik, Georg-August-Universit\"at G\"ottingen, Friedrich-Hund-Platz 1, 37077 G\"ottingen \\
\email{aboesch@astro.physik.uni-goettingen.de}}
\date{Received 18 April 2015 / Accepted 10 July 2015}

 
\abstract
{A discharge of nitrogen gas, as created in a microwave-induced plasma, exhibits a very dense molecular emission line spectrum. Emission spectra of this kind could serve as wavelength calibrators for high-resolution astrophysical spectrographs in the near-infrared, where only very few calibration sources are currently available.}
{The compilation of a spectral line list and the characterization of line intensities and line density belong to the initial steps when investigating the feasibility of potential wavelength calibration sources. Although the molecular nitrogen spectrum was extensively studied in the past, to our knowledge, no line list exists that covers a continuous range of several thousand wavenumbers in the near-infrared.}
{We recorded three high-resolution ($\Delta \tilde{\nu} = 0.018$\,cm$^{-1}$) spectra of a nitrogen gas discharge operated at different microwave powers. The nitrogen gas is kept inside a sealed glass cell at a pressure of 2\,mbar. The emission lines in the spectra were fitted by a superposition of Gaussian profiles to determine their position, relative intensity, and width. The line parameters were corrected for an absolute wavelength scale, instrumental line broadening, and intensity modulation. Molecular and atomic transitions of nitrogen were identified with available line positions from the literature.}
{We report line lists with more than 40\,000 emission lines in the spectral range $4500-11000$\,cm$^{-1}$ ($0.9-2.2$\,$\mu$m) \thanks{The line lists and spectra are available in electronic form at the CDS via anonymous ftp to cdsarc.u-strasbg.fr (130.79.128.5) or via http://cdsarc.u-strasbg.fr/viz-bin/qcat?J/A+A/vol/page}. The spectra exhibit emission lines over the complete spectral range under investigation with about $350-1300$ lines per 100\,cm$^{-1}$. Depending on the microwave power, a fraction of $35\% - 55\%$ of all lines are blended. The total dynamic range of the detected lines covers about four orders of magnitude.}
{Line density and peak intensities qualify the recorded discharge as a useful wavelength calibrator, and the line list provides an empirical reference for nitrogen spectra in the near-infrared.}


\titlerunning{Spectral line lists of a nitrogen gas discharge for wavelength calibration}
\maketitle

\section{Introduction}
\label{sec: Introduction}

High-resolution near-infrared spectroscopy is an increasingly important tool in astrophysics. A major driver for this development is research on cool low-mass stars, such as M dwarfs, which emit most of their photons at wavelengths above 1\,$\mu$m. For example, M dwarfs are of great importance in the search for low-mass extrasolar planets and are in the focus of upcoming observing programs \citep{Reiners2010, Quirrenbach2014}. A key element of precision spectroscopy in astronomy is a reliable wavelength calibration, which assigns the wavelength scale to the pixels of the spectrograph's detector.

Currently, the two most frequently used wavelength calibration techniques are absorption gas cells (e.g., filled with iodine) or hollow cathode lamps \citep[e.g.,][]{Marcy1992, Baranne1996}. Emission lamps with atomic line spectra have been used at optical wavelengths with great sucess, but they feature fewer lines toward longer wavelengths. The line list for commercial thorium-argon hollow cathode lamps contains about 2400 spectral lines suitable for wavelength calibration in the range $900-4500$\,nm \citep{Kerber2008}. Absorption cells with improved gas mixtures, and new techniques, such as laser frequency combs and Fabry-Perot etalons, are under development \citep{Seemann2014, Wilken2012, Schaefer2012}. Nevertheless, molecular emission spectra, e.g., from N$_2$ or CN, can be a viable alternative for wavelength calibration \citep{Boesch2014}: a discharge of nitrogen gas exhibits a very dense line spectrum in the near-infrared, in contrast to hollow cathode lamps; the frequency of all lines are defined by physical laws not supposed to change over time, in contrast to Fabry-Perot etalons; and the equipment is relatively cheap, in contrast to laser frequency combs.

For the investigation of new calibration sources, the compilation of a spectral line list and the characterization of line intensities and distribution are initial steps. Although the spectrum of molecular nitrogen has been extensively studied in the past, a line list based on high-resolution observations over a wide range in the near-infrared, to our knowledge, does not exist. 

An extensive compilation of spectroscopic data on the molecule N$_2$ (and its ions N$_2^-$, N$_2^+$ and N$_2^{2+}$) was presented by \citet{Lofthus1977}. Their review discusses each electronic band system, provides extensive references to previous research, and can be used to identify band systems expected in a given wavelength range. For N$_2$, the First Positive System B$^3\Pi_\text{g}$-A$^3\Sigma_\text{u}^+$ ($478-2531$\,nm) is the most prominent band system and appears in most types of discharges according to \citet{Lofthus1977}. Furthermore, the Meinel System A$^2\Pi_\text{u}$-$X ^2\Sigma_\text{g}^+$ of N$_2^+$ has transitions in the near-infrared ($550-1770$\,nm). More recent reports on spectroscopic data for molecular nitrogen in general can, e.g., be found in \citet{Laher1991}, \citet{Gilmore1992} and \citet{Laux1992}.

The First Positive System was specifically analyzed by \citet{Effantin1979} and \citet{Roux1983}. The paper by \citet{Effantin1979} covers the vibrational bands (0-0), (1-0) and (2-0) and includes line lists of observed and calculated wavenumbers. The study of the First Positive System was later extended to 33 bands by \citet{Roux1983}, but they did not publish their line list. Our efforts to retrieve the list from the journal or the institute where the research was conducted were not successful. For the Meinel System of $^{14}$N$_2^+$, line lists of observed and calculated wavenumbers for the (0-0), (0-1), (1-0) and (1-2) bands are included in the work by \citet{Ferguson1992}.

In this article, we present the characterization of a nitrogen gas discharge observed in the near-infrared with a Fourier transform spectrometer (FTS). Three spectra were recorded with the discharge operated at different microwave (MW) powers. We describe the experimental procedure and the data reduction process in detail. We fit all emission lines above a minimum flux threshold and produce a list including line position, line intensity, and line width. The identification of all transitions based on theoretical calculations is beyond the scope of this work, but we include assignments of molecular nitrogen transitions from the (0-0) band of the First Positive System and from four bands of the Meinel System, in addition to atomic nitrogen lines. We analyze properties of line intensity and line density with a focus on the spectrum's usefulness as a wavelength calibrator in astrophysical spectroscopy. For the discharge operated at different MW powers, we investigate the change in line width of detected lines and in line intensity of identified atomic nitrogen.

\section{Experiment}
\label{sec: experiment}

\begin{figure}
\resizebox{\hsize}{!}{\includegraphics{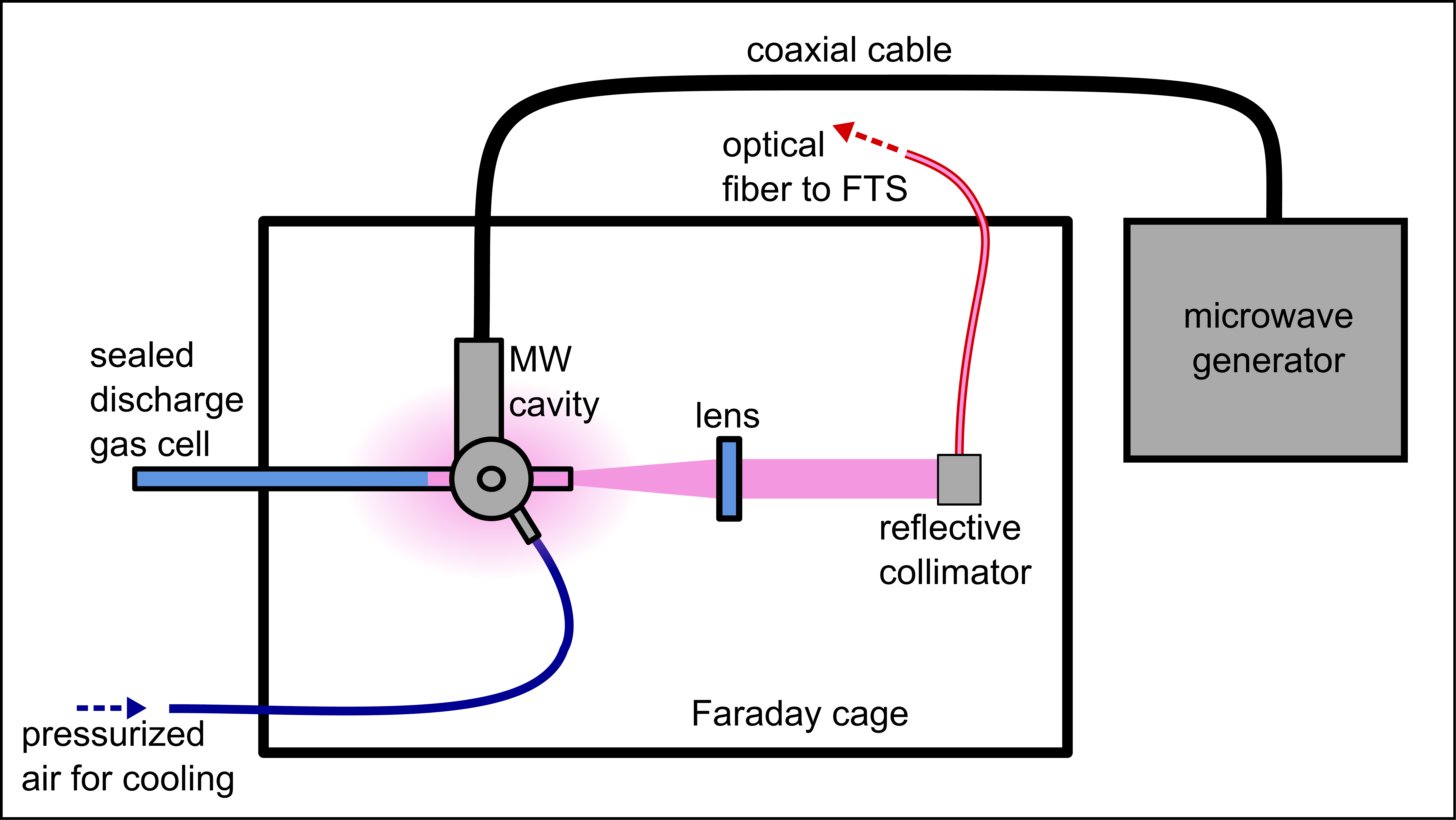}}
\caption{The experimental setup. See main text for explanation (Sect.~\ref{sec: experiment}).}
\label{fig: experiment}
\end{figure}

The experimental setup is similar to that described in \cite{Boesch2014}. Fig.~\ref{fig: experiment} shows a schematic drawing. Here, we summarize the general functionality of the experiment for completeness and highlight the details that are important for the measurements presented here.

We use an Evenson-type microwave cavity (from Opthos Instruments\footnote{Commercial products are identified in this document to specify the experimental procedure adequately. This identification is not intended to imply recommendation or endorsement, nor is it intended to imply that the products identified are necessarily the best available for the purpose.}) and a solid-state microwave generator (GMS 200W from Sairem) to create a glowing plasma inside a sealed discharge gas cell. The cell has a diameter of 12\,mm, a length of about 20\,cm, and is made of quartz glass. A plane quartz window is melted to the front end with a tilt of 5$^\circ$. The cell, filled with 2\,mbar of high-purity dry nitrogen gas, was purchased from Sacher Lasertechnik. A sealed gas cell enables us to make repeated measurements with the same gas properties and allows for a compact experimental design.

The emitted light passes through a convex lens (Thorlabs LA1255) and is focused by a reflective collimator (Thorlabs RC08SMA-P01) onto an optical fiber (Thorlabs FG550LEC). The fiber feeds the lights into an FTS (Bruker IFS 125HR). On February 14, 2015, we recorded three high-resolution spectra ($\Delta \tilde{\nu} = 0.018$\,cm$^{-1}$; spectral resolving power of $R=555556$ at $\lambda = 1\,\mu$m) of the nitrogen gas discharge. For each spectrum, the discharge was operated at a different MW power. The forward MW power set at the generator was 25\,W, 50\,W, or 100\,W for the three measurements. The chosen resolution allowed us to record a symmetric interferogram that minimizes phase corrections errors. For each of the three measurement runs, the final interferogram is an average of 300 scans. The parameters of the FTS are given in Table~\ref{tab: FTS}.

For the determination of the absolute wavenumber scale, we additionally recorded spectra of a sealed discharge cell filled with 2\,mbar of argon gas. The argon gas cell simply replaced the cell filled with nitrogen for these measurements.

Flat-field spectra were measured using a laboratory halogen lamp (Ocean Optics HL-2000-FHSA-HP) to correct spectral line intensities for the transmittance of optical elements and the detector response. For the flat-field measurements, the gas cell was removed from the experimental setup and the halogen lamp was placed behind the MW cavity.

\begin{figure*}
\sidecaption
\includegraphics[width=12cm]{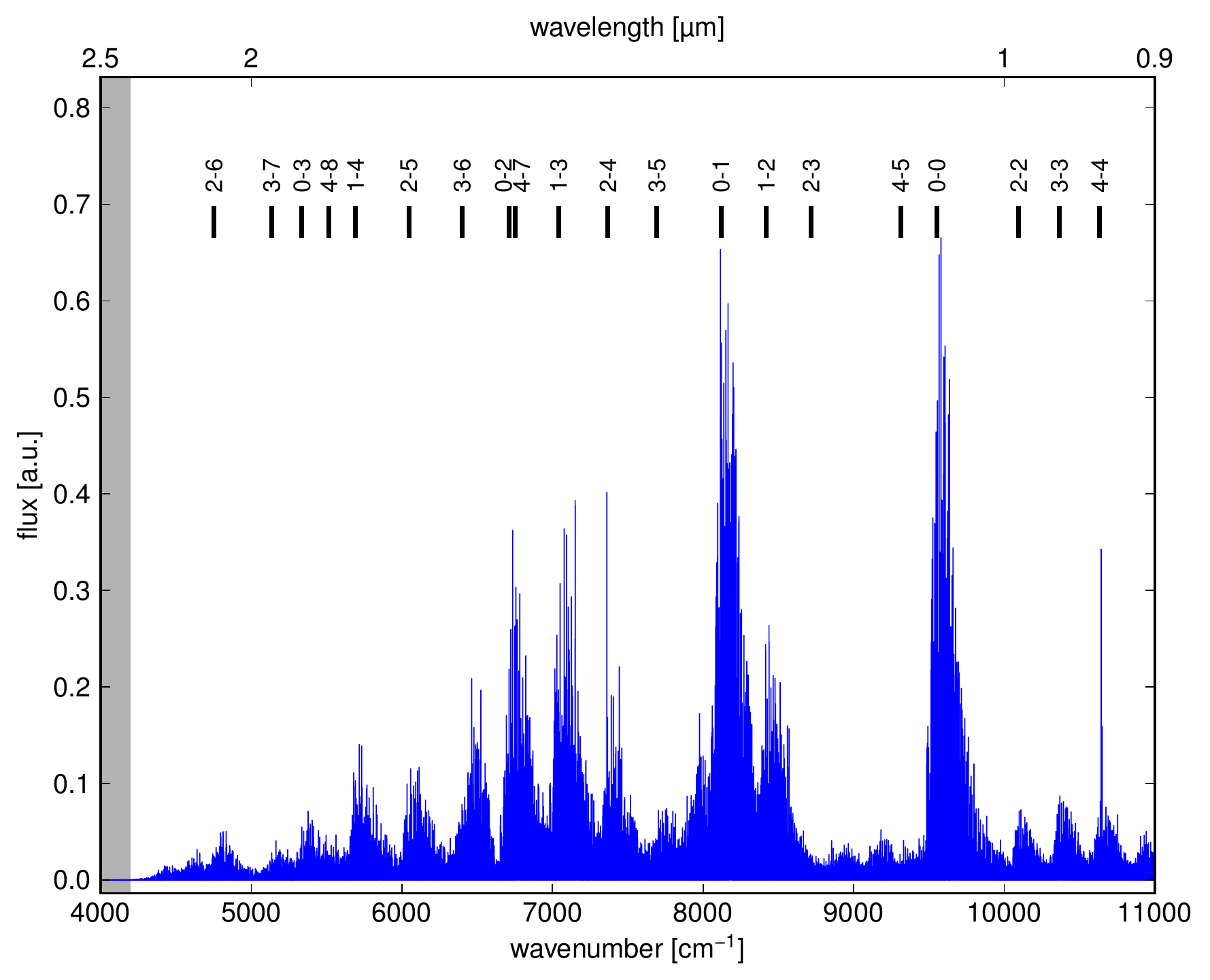}
\caption{Raw emission spectrum of the nitrogen gas discharge operated at a MW power of 50\,W. Strong vibrational bands from the First Positive System of $^{14}$N$_2$ are labeled ($v'-v''$) for orientation using band origin wavenumbers as given by \citet{Gilmore1992}. The spectral range below 4200\,cm$^{-1}$ (gray box) does not contain any emission lines and was used to determine the absolute noise level in the spectrum.}
\label{fig: spectrum}
\end{figure*}

Figure~\ref{fig: spectrum} shows the raw spectrum of the nitrogen gas cell, covering the wavenumber range $4000-11000$\,cm$^{-1}$ ($0.9-2.5$\,$\mu$m). Intensities are given in arbitrary units (a.u.) as obtained from the Fourier transform. The transmittance of the optical fiber and the response curve of the detector set the limits of the usable spectral range. No signal above noise was detected at wavenumbers below 4200\,cm$^{-1}$ (above $2.38\,\mu$m). This part of the spectrum is indicated in Fig.~\ref{fig: spectrum} with a shadowed box and was used for determining the absolute noise level (see Sect.~\ref{sec: noise}). The spectra are available in the online material.

\begin{table}
\caption{Settings of the FTS for each of the three measurements with different MW powers.}
\label{tab: FTS}
\centering
\begin{tabular}{l l}
\hline \hline
resolution & 0.018\,cm$^{-1}$ ($\times 0.944$\tablefootmark{a})\\
scans      & 300 ($\approx$ 6\,h 45\,min total scanning time) \\
aperture   & 1.3\,mm \\
detector   & InSb (liquid nitrogen cooled) \\
beamsplitter & CaF$_2$ \\
pressure inside FTS & $0.10-0.15$\,mbar \\
zerofilling factor & 4 \\
\hline
\end{tabular}
\tablefoot{
\tablefoottext{a}{The applied apodization (Norton-Beer medium) modifies the resolution set in the spectrometer control software by this factor.}
}
\end{table}

\section{Data Analysis}

This section explains the steps in the data reduction to obtain the parameters of the spectral emission lines. First, we determine the absolute noise level in the spectrum. Then, the emission lines are located and fitted. The line parameters are corrected for absolute wavenumber scale, instrumental broadening, and intensity modulation. Finally, spectral lines are identified with available line positions from the literature.

\subsection{Noise level}
\label{sec: noise}
The knowledge of the noise in the spectrum is important to properly weight the data points and derive meaningful errors for the line parameters through error propagation. Under the assumption of white noise in the interferogram, the noise in a spectrum obtained from an FTS is uniformly distributed \citep{Davis2001, Voigtman1987}. The premise of white noise is reasonably valid for the dominance of detector or photon noise. The determination of the noise in a single spectrum without a sufficient continuum poses a practical difficulty. In our case, the spectrum has a baseline without any spectral lines only below 4200\,cm$^{-1}$ (marked with a gray box in Fig.~\ref{fig: spectrum}). We use this wavenumber region to calculate the noise.

Additionally, the baseline of the spectrum has a small offset from zero, which is best visible in the region without spectral features. The offset can be explained with an uncertainty in the first component of the Fourier transform, i.e., the intensity of the interferogram at zero path difference. Because this value is wavelength independent, we subtract the median flux from the wavenumber region $4000-4200$\,cm$^{-1}$ from the whole spectrum before further analysis.

\begin{figure}
\resizebox{\hsize}{!}{\includegraphics{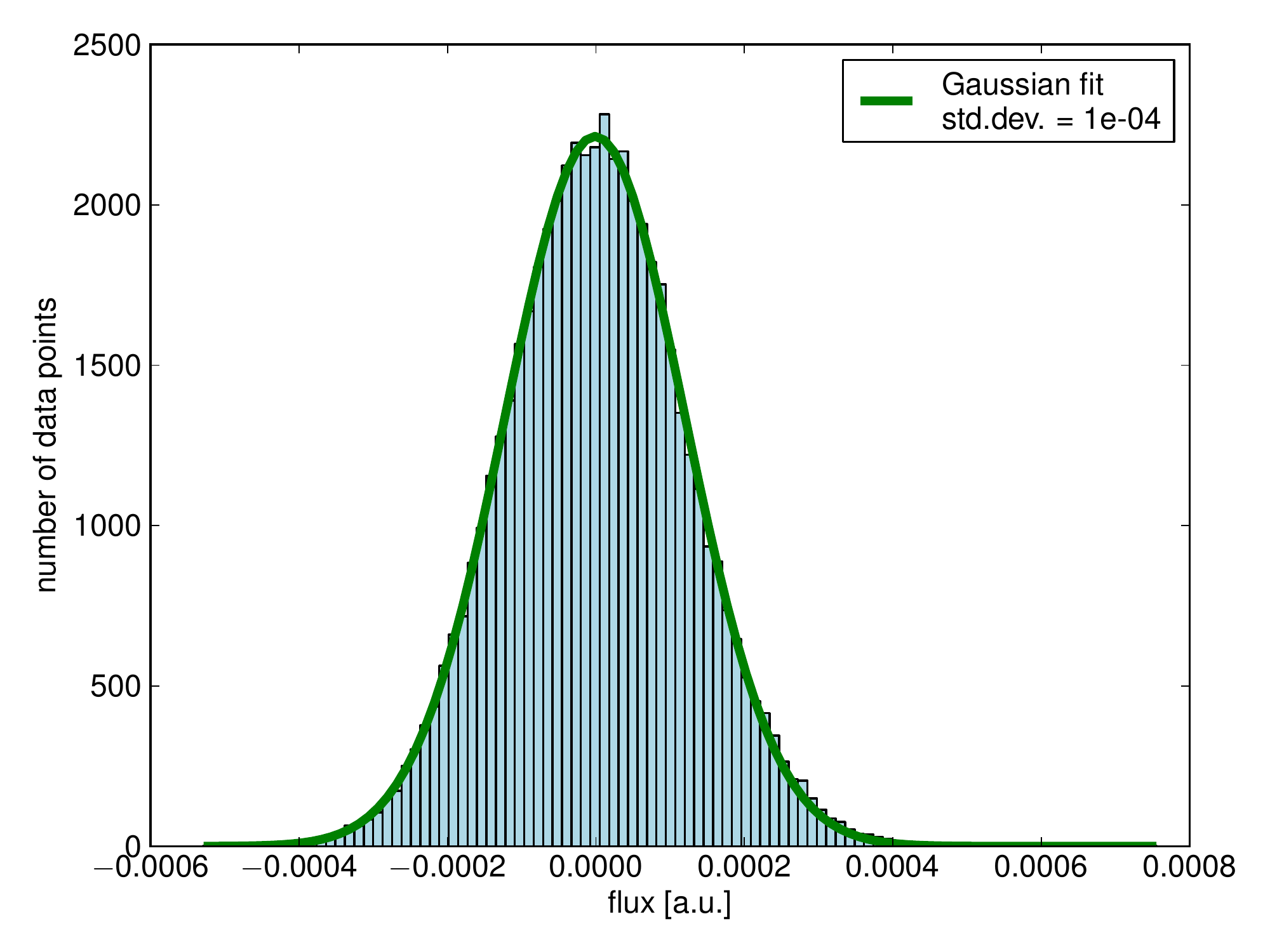}}
\caption{Histogram of the flux values in the spectral region $4000-4200$\,cm$^{-1}$ from the measurement of the nitrogen discharge operated at a MW power of 50\,W. A Gaussian profile is fitted to the histogram.}
\label{fig: noise}
\end{figure}

Then, we arrange the flux values from the wavenumber region $4000-4200$\,cm$^{-1}$ in a histogram and fit a Gaussian function to it describing the distribution very well; Fig.~\ref{fig: noise} shows this exemplary for the measurement of the discharge operated at a MW power of 50\,W. We adopt the standard deviation of the Gaussian profile as the absolute noise level. The first two rows in Table~\ref{tab: dataanalysis} list the subtracted offset values and the noise levels for the three measurements.

\begin{table}
\caption{Baseline offset, absolute noise level, minimum peak intensity threshold, and number of located emission peaks for the three measurements.}
\label{tab: dataanalysis}
\centering
\begin{tabular}{l|ccc}
\hline \hline
                              & 25\,W   & 50\,W   & 100\,W \\
\hline
baseline offset [a.u.]        & 0.00009 & 0.0001 & 0.0001       \\
noise level [a.u.]            & 0.0001  & 0.0001 & 0.0001       \\
minimum peak intensity [a.u.] & 0.001   & 0.001  & 0.001       \\
number of peaks found         & 40408   & 51776  & 58274        \\
\hline
\end{tabular}
\end{table}

\subsection{Peak finding and peak fitting}
\label{sec: findfit}
We locate the emission peaks in the spectrum, together with the intensity at these points, using the zero-crossings in the spectrum's first derivative. The derivative is calculated at every data point after the entire spectrum has been interpolated with a cubic spline. To be classified as a peak, a data point must fulfill the following three criteria: (i) a zero-crossing in the first derivative, (ii) an intensity in the spectrum above a threshold value of ten times the absolute noise level, and (iii) a maximum intensity within one resolution element. Experience shows that the adopted minimum intensity threshold value is a good compromise between considering as many peaks as possible and satisfactory fit results. The minimum peak intensity (ten times the absolute noise level) and the number of located peaks are given in Table~\ref{tab: dataanalysis}. The peak search and the final line list cover the wavenumber range $4500-11000$\,cm$^{-1}$.

The first derivative's zero-crossing technique delivers estimates of the lines' positions and intensities. The estimates serve as initial values for the subsequent fitting process, which is needed to determine accurate line parameters. The fitting procedure employs a weighted least-squares minimization (package \textit{mpfit} for the \textit{Python} programming language\footnote{http://cars.uchicago.edu/software/python/mpfit.html}). Multiple emission lines are simultaneously fitted with a superposition of Gaussian profiles. The combined fit of several spectral lines is important  to account for overlapping lines. 

The fit with a Gaussian profile gives the line position $\tilde{\nu}_0$, line intensity $I$, and line width $\sigma_\text{G}$ (standard deviation). We extract the uncertainties for the parameters from the covariance matrix of the fitting routine. The full width at half maximum (FWHM) of a line can be calculated using $\text{FWHM} = 2\sigma_\text{G}\sqrt{2\ln(2)}$. We also tested a Voigt profile for the fit function, but did not obtain better values for the reduced $\chi^2$ compared to the Gauss profile. We rather observed that the Voigt profile is strongly influenced by (blended) low-intensity lines near the noise level because the overall shape of the Lorentzian profile is more sensitive to data points in the wings of the lines.

\subsection{Correction of absolute wavenumber scale}
\label{sec: kfactor}
One advantage of Fourier transform spectroscopy is the high relative precision of the wavenumber scale. The absolute wavenumber scale as given by the internal reference laser is, however, subject to changes due to misalignments and environmental changes inside the FTS \citep[see, e.g.,][chap. 2.6]{Griffiths2007}. The resultant shift in the wavenumber scale is linear in wavenumber and can be corrected using known positions of calibration lines. In theory, one reference line with a well-known line position is sufficient, but in practice a set of lines is used. The correction factor $\kappa_\text{eff}$ is then the weighted average of $\kappa = (\tilde{\nu}_{0\text{, ref}} / \tilde{\nu}_{0\text{, measured}})-1$ for each reference line, where $\tilde{\nu}_{0\text{, ref}}$ is the line position from the literature and $\tilde{\nu}_{0\text{, measured}}$ is the line position in the recorded spectrum. The following calibration is applied to all detected spectral lines:
\begin{align}
\tilde{\nu}_{0\text{, c}} = \tilde{\nu}_0\left(1+\kappa_\text{eff} \right)  ,
\end{align}
where $\tilde{\nu}_{0\text{, c}}$ is the calibrated wavenumber and $\tilde{\nu}_0$ is the uncalibrated wavenumber.

\begin{figure}
\resizebox{\hsize}{!}{\includegraphics{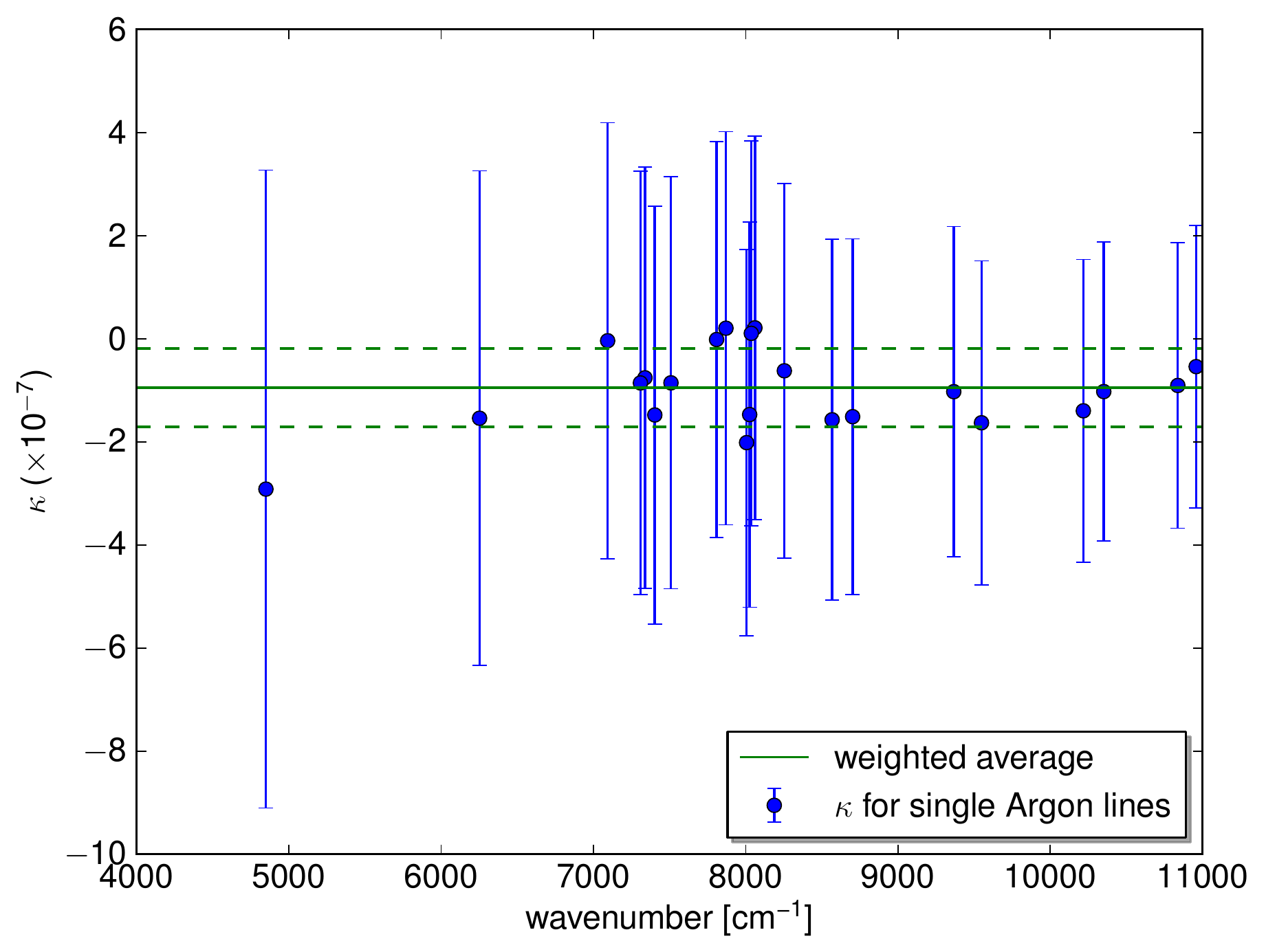}}
\caption{The correction factor $\kappa$ for the absolute wavenumber scale of the FTS as determined for 22 atomic argon lines. The horizontal line indicates the weighted average (solid line) with its error margin (dashed lines). Shown are the data for the argon spectrum recorded before the measurements of the nitrogen spectra.}
\label{fig: k-factor}
\end{figure}

To determine the absolute wavenumber scale, we compared positions of Ar I lines with data from the NIST database \citep{NIST_ASD}. We recorded spectra of an argon gas discharge before and after the measurements of the nitrogen spectra. The line positions utilized from the NIST database are given with an accuracy of three decimal places. Additionally, taking  information from the original reference of the argon transitions \citep{Norlen1973} into account, we conservatively assume an uncertainty of 0.003\,cm$^{-1}$. The errorbars on $\kappa$ for the individual argon lines (as shown in Fig.~\ref{fig: k-factor}) are calculated by error propagation of the uncertainties of the position of the reference line and the line's position in our spectra as determined from a line fit. Using 22 argon lines, we calculated  correction factors of $\kappa_\text{eff, before} = (-0.9 \pm 0.8) \cdot 10^{-7}$ and $\kappa_\text{eff, after} = (-0.8 \pm 0.8) \cdot 10^{-7}$ from the observations taken before and after the measurements of the nitrogen spectra, respectively. Combining the two results, we accept a value of $\kappa_\text{eff} = (-0.85 \pm 0.85) \cdot 10^{-7}$. This relates to an uncertainty in the corrected wavenumber scale of $(0.85\cdot 10^{-7}) \tilde{\nu}_0$, namely about 0.00038\,cm$^{-1}$ at 4500\,cm$^{-1}$ and about 0.00094\,cm$^{-1}$ at 11000\,cm$^{-1}$.

\subsection{Correction for instrumental resolution}
\label{sec: corrlinewidth}
The shape of a recorded spectral line is the convolution of its true shape and the instrumental line spread (ILS) function. We assume the ILS function to be of Gaussian shape. The convolution of two Gaussian functions results in another Gaussian with a variance that is equal to the sum of the variances of the original functions. Under the assumption of a Gaussian ILS, we can estimate the FWHM before the instrumental broadening by applying
\begin{align}
\text{FWHM}_\text{c} = \sqrt{\text{FWHM}^2-\text{FWHM}_\text{instr}^2} ,
\end{align}
where $\text{FWHM}$ and $\text{FWHM}_\text{c}$ are the uncorrected and corrected values, respectively. $\text{The FWHM}_\text{instr}=(0.018\cdot 0.944)$\,cm$^{-1}$ is the theoretical FWHM of the ILS function (compare with the resolution given in Table~\ref{tab: FTS}). Using FTS observations of a frequency-locked diode laser with a line width of $\approx$100\,kHz, we found that the true ILS tends to be slightly broader than the theoretical value, but the difference is less than 10\%.

\subsection{Correction of line intensities}
\label{sec: corrff}
\begin{figure}
\resizebox{\hsize}{!}{\includegraphics{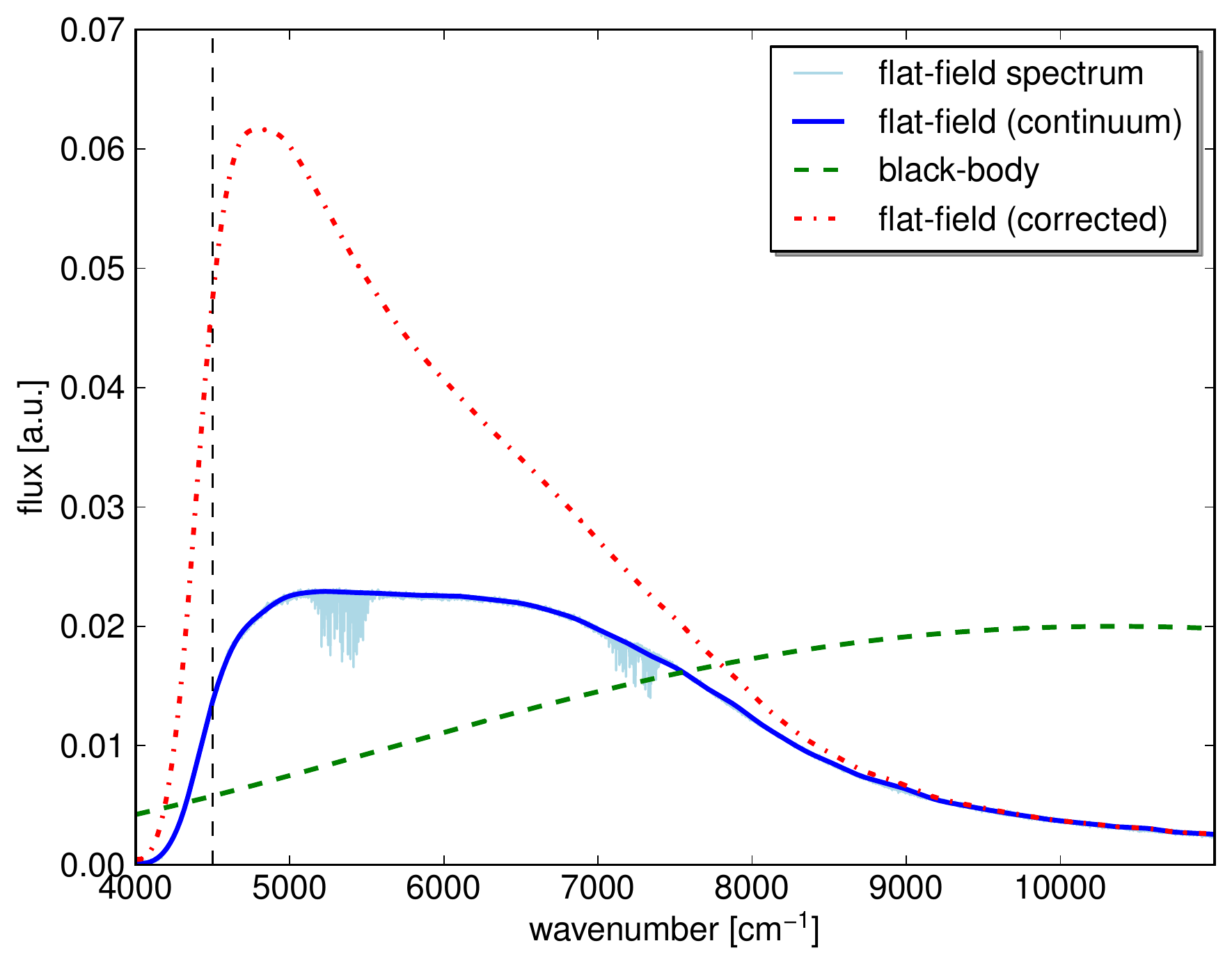}}
\caption{The spectrum of a halogen lamp is used as a flat-field (spectrum shown in light blue with absorption bands from atmospheric water). The dark blue line describes the continuum of this spectrum. The green dashed curve shows the theoretical normalized curve of a blackbody at the halogen lamp's bulb temperature of 3000\,K (scaled by a factor of 1/50 for visualization). The red dot-dashed curve is the flat-field continuum divided by the theoretical normalized blackbody spectrum. The raw intensities of the spectral lines are corrected by the latter curve. The vertical dashed line indicates 4500\,cm$^{-1}$.}
\label{fig: flat-field}
\end{figure}

The intensity of the spectrum is modulated because of the transmittance of the optical elements in the experimental setup and the response curve of the detector. The true intensities of the emission lines can be retrieved through comparison with a spectrum of known shape, i.e., a flat-field spectrum. The spectrum of a laboratory halogen lamp was recorded before and after the measurements of the nitrogen gas cell, and the two spectra were averaged to give one flat-field spectrum (see Fig.~\ref{fig: flat-field}). Theoretically, the halogen lamp produces a blackbody spectrum with a maximum intensity at 969\,nm, which corresponds to the bulb temperature of 3000\,K, as referenced by the manufacturer. The spectral shape of the blackbody is not constant over the whole wavenumber range under investigation (green dashed curve in Fig.~\ref{fig: flat-field}). We correct the flat-field spectrum's continuum by the theoretical curve and use the result for the intensity correction: the line intensities $I$ derived from the spectra are divided by the intensity of the corrected flat-field curve at their position (red dash-dotted line in Fig.~\ref{fig: flat-field}). As result, we obtain the flat-field corrected peak intensities $I_\text{c}$.

The degree of knowledge of the lamp's true spectrum and the deviation from the assumed theoretical blackbody limit the accuracy in the flat-fielding process. The spectral output of a similar lamp (Ocean Optics HL-2000-FHSA) was measured by the manufacturer with respect to a calibrated system at wavenumbers above 6250\,cm$^{-1}$ (up to 1.6\,$\mu$m; private communication). The maximum deviation of the measured spectral output from a theoretical blackbody with the temperature as stated in the specifications was less than 7\%. Taking further uncertainties in the flat-fielding process into account, we assume that the error on $I_\text{c}$ is around 10\% and therefore small compared to the intensity modulation by the detector response, which can be inferred from Fig.~\ref{fig: flat-field}.

\subsection{Atomic line identification}

All emission peaks in the spectrum were checked against line lists from the atomic line database from NIST \citep{NIST_ASD} for N I, N II, O I, and Ar I. We checked for oxygen and argon because they are the most abundant species in the atmosphere after nitrogen, and would be obvious contaminators, if the cell was not filled and sealed properly. We only used lines with observed wavelength values in the NIST database. For a match, we require the distance between the recorded line and the database to be less than the line's half width at half maximum ($\text{HWHM}_\text{c} = \text{FWHM}_\text{c}/2$). The probability that the position of a line from the literature matches a line in our recorded spectrum by chance is 20\%, 30\% and 38\% for the measurements with MW power of 25\,W, 50\,W and 100\,W, respectively, because of the many densely-spaced spectral lines.

Although up to 35 oxygen lines (53\% of the lines listed in the NIST database) and 35 argon lines (57\%) fulfill the above mentioned HWHM$_\text{c}$ criterion, we judge them to be misidentified in the crowded spectrum. We come to this assessment, because the determined FWHM$_\text{c}$ of the lines do not follow the expected linear trend with wavenumber (see Sect.~\ref{sec: MWPower} and compare with Fig.~\ref{fig: FWHM}) and the peak intensities $I_\text{c}$ do not change by the same amount when the MW power is altered (compare with Fig.~\ref{fig: intenschange_NI}). The absence of lines from oxygen and argon indicates that no leakage of air into the gas cell had occurred.

For nitrogen, only lines from the neutral atom could be identified with confidence. This is not surprising because the degree of ionization is low in this kind of electrodeless discharge lamp \citep{Thorne1999}.

\subsection{Identification of molecular nitrogen lines}

We use the spectral line lists with branch designations for molecular nitrogen transitions in the wavenumber range of our spectra from \citet{Effantin1979} and \citet{Ferguson1992}. As for the atomic line identification, we require that the distance to the literature value is less than the line's half width at half maximum.

The work by \citet{Effantin1979} provides line lists for the \mbox{$(0-0)$} band of the B$^3\Pi_\text{g}$-A$^3\Sigma_\text{u}^+$ system of $^{14}$N$_2$ and $^{15}$N$_2$. The lists consist of measured wavenumbers for 21 branches and calculated wavenumbers for 27 branches. The observed and calculated wavenumbers differ by less than 0.01\,cm$^{-1}$, where both values are given. \citet{Effantin1979} did not observe lines from the six branches designated R$_{13}$, Q$_{13}$, P$_{13}$, R$_{31}$, Q$_{31}$,  and P$_{31}$. Our nitrogen gas contains the natural ratio of isotopes ($^{14}$N$_2:^{15}$N$_2=99.636:0.364$; \citealt{Lide2006}) and therefore only very faint lines of $^{15}$N$_2$ are expected, if any at all. We do not include them in the analysis. For $^{14}$N$_2$, we can identify up to 80\% of the lines from the literature, including some from the above mentioned branches that were not observed by \citet{Effantin1979}. 

Furthermore, we identified up to 89\% of the lines from four vibrational bands of the Meinel System of $^{14}$N$_2^+$ as listed in \citet{Ferguson1992}. The four bands have their origins at 9016\,cm$^{-1}$ \mbox{$(0-0)$}, 6841\,cm$^{-1}$ \mbox{$(0-1)$}, 10889\,cm$^{-1}$ \mbox{$(1-0),$} and 6572\,cm$^{-1}$ \mbox{$(1-2)$}.

The HITRAN2012 database \citep{Rothman2013} lists additional 584 quadrupol transitions in the electronic ground state of $^{14}$N$_2$ (in the wavenumber range $4500-11000$\,cm$^{-1}$). The number of matches in line position (17\%, 26\%, and 34\% for our three measurements with MW power of 25\,W, 50\,W, and 100\,W, respectively) is below the probability for a random match for each spectrum, as stated in the previous section.

\subsection{Examples of line fitting}
\label{sec: examples}

\begin{figure}
\resizebox{\hsize}{!}{\includegraphics{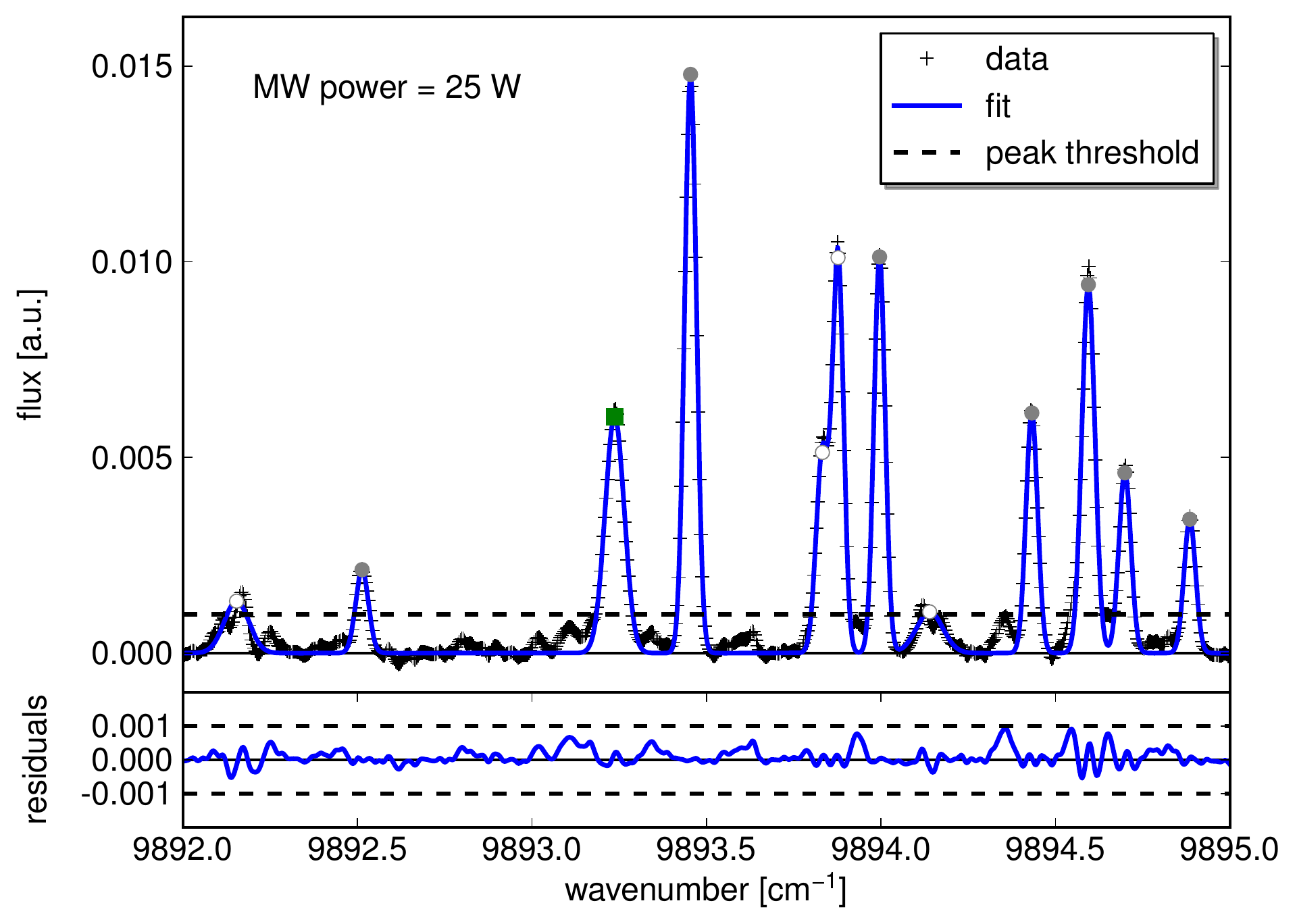}}
\resizebox{\hsize}{!}{\includegraphics{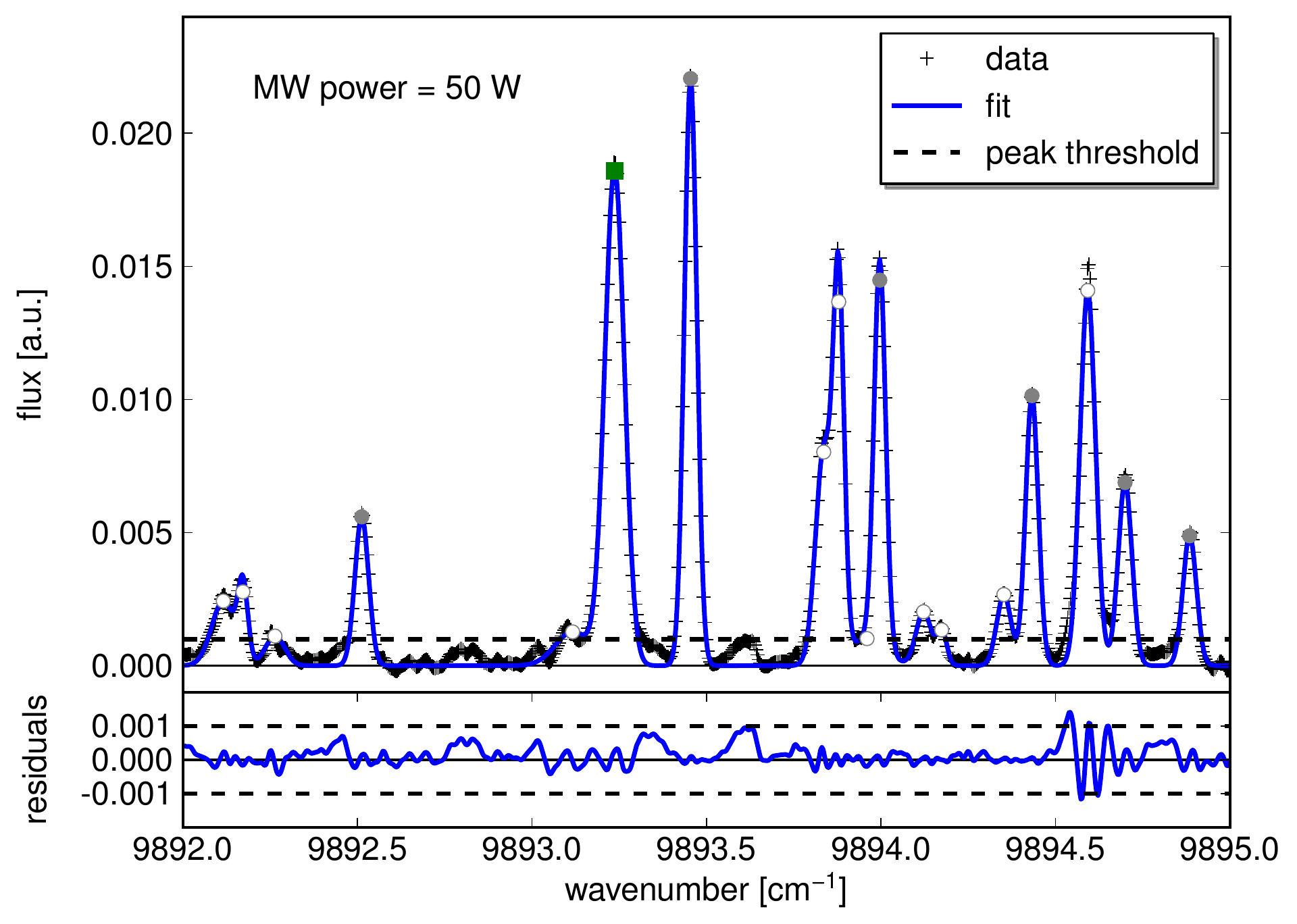}}
\resizebox{\hsize}{!}{\includegraphics{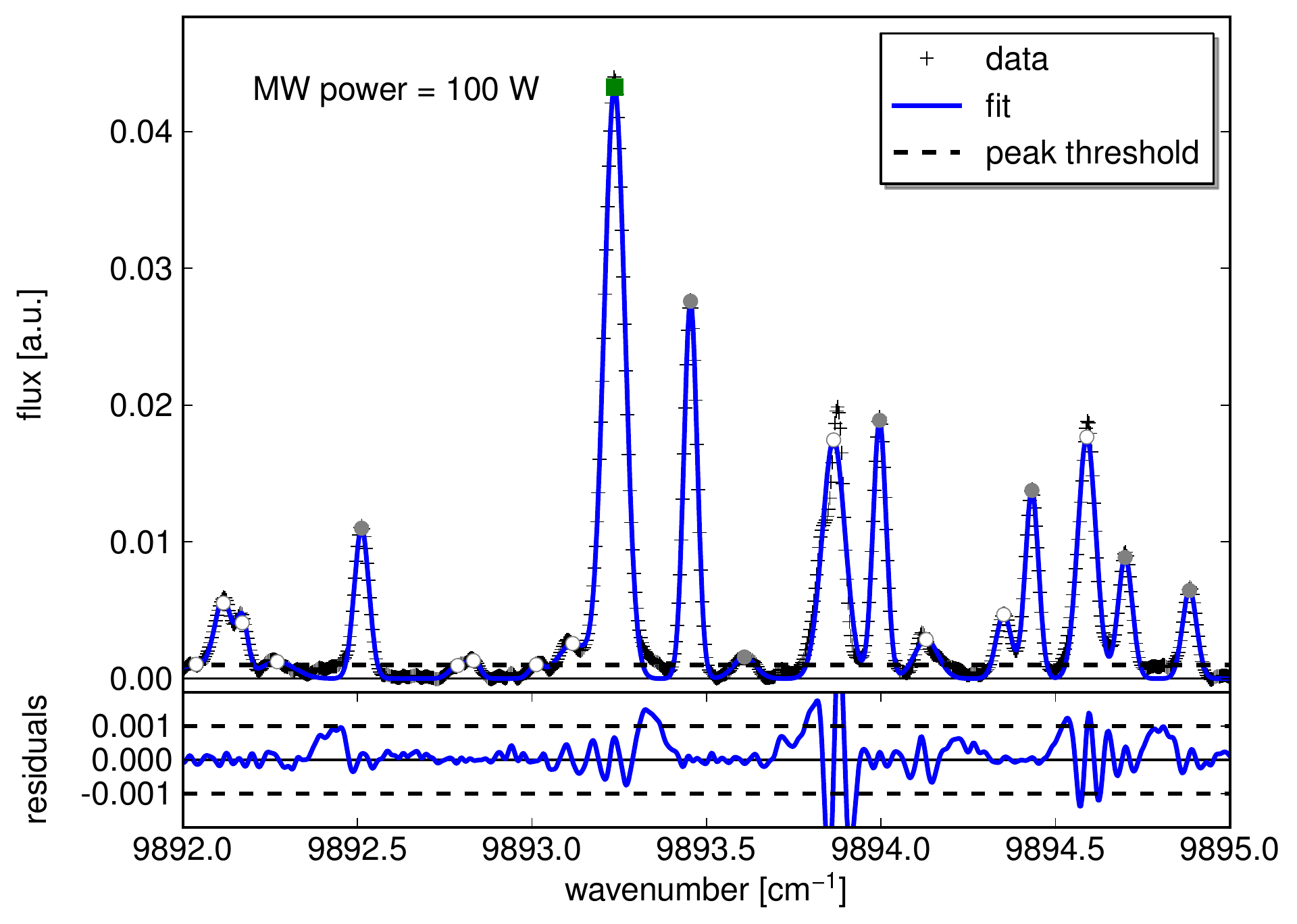}}
\caption{The recorded spectra between 9892 and 9895\,cm$^{-1}$. From top to bottom: discharge operated at MW power of 25, 50, and 100\,W. The blue line is the fit composed of a superposition of Gauss profiles to the peaks exceeding the minimum peak intensity threshold (dashed horizontal line). A green square indicates an identified N I line, while a gray circle marks an unidentified line. See main text for more information (Sect.~\ref{sec: examples}). Note that the ordinate axis is not the same scale for each plot.}
\label{fig: identspectrum_all}
\end{figure}

We present in Fig.~\ref{fig: identspectrum_all} a part of the three spectra taken at different MW powers. Each of the three panels shows the same spectral interval ($9892-9895$\,cm$^{-1}$). The black crosses are the data points of the spectrum and the blue curve represents the fit to the emission lines. The black dashed horizontal bar indicates the minimum peak intensity (peak threshold). The peaks are marked at their line center and maximum line flux, as determined by the fit. A spectral line is marked with a filled symbol if no noticeable problems during the analysis process occurred, while an empty symbol indicates a line with one or more flags (see Sect.~\ref{sec: linelist}). A green square identifies an N I line and a gray circle marks an unidentified line. The lower part of each panel shows the fit residuals (blue curve).

We observe in Fig.~\ref{fig: identspectrum_all} that line intensities grow with increasing microwave power. At the same time, the noise level remains constant (compare Table~\ref{tab: dataanalysis}), and therefore more data points exceed the peak threshold and are included in the line fit.

\section{Results}
\label{sec: results}

In the following section, we present the spectral line lists created from the measurements of the nitrogen gas discharge and some properties of the line parameters. The analysis of line intensities and line density focuses on the usability of this light source for wavelength calibration of astrophysical spectrographs. In Sect.~\ref{sec: MWPower} we demonstrate the influence of the MW power level on the line width of detected lines in general, and, specifically, on the peak intensities of identified N~I lines.

\subsection{Spectral line list}
\label{sec: linelist}
We created separate line lists for the three measurements that include all detected emission peaks with their parameters in the wavenumber range $4500-11000$\,cm$^{-1}$ ($0.9-2.2$\,$\mu$m). The line lists are available in the online material as simple text files consisting of 15 columns. Table~\ref{tab: listexplanation} gives a description of the content in each column. The stated errors of the positions $\sigma(\tilde{\nu}_0)$, line intensities $\sigma(I),$ and line widths $\sigma(\sigma_\text{G})$ are relative errors corresponding to the line fit in the spectrum. They are taken from the covariance matrix of the weighted least-squares algorithm (see Sect.~\ref{sec: findfit}). The error on the absolute line position is less than 0.00094\,cm$^{-1}$ as derived in Sect.~\ref{sec: kfactor}. The error on the FWHM is dominated by the uncertainty on the instrumental line profile as described in Sect.~\ref{sec: corrlinewidth}. The maximum error on the relative intensities due to uncertainties in the flat-fielding procedure is estimated to be around 10\% (see Sect.~\ref{sec: corrff}).

Column 14 contains information on problems, which were noticed during the data reduction process and could have an influence on the determined line parameters. The information is given in the form of up to five letters that represent different cases as explained in Table~\ref{tab: listexplanation}. We call these letters flags throughout this article.

The number of spectral lines in the three spectra are summarized in Table~\ref{tab: results}. A total of 40408, 51776, and 58274 lines were detected and fitted for MW powers of 25\,W, 50\,W, and 100\,W, respectively. This confirms the trend we have already seen in Fig.~\ref{fig: identspectrum_all}: the total number of detected lines grows with increasing MW power. Table~\ref{tab: results} also gives the number of identified lines for the different species. More lines from the literature are identified when the discharge is operated with a higher MW power because both the intensity of the lines increases (more fainter lines are included) and the lines become broader (makes a match with line positions from the literature more likely).

The majority of spectral lines remains unidentified. Spectral modeling and calculated transition wavenumbers are needed to confirm that the unidentified lines in fact originate from nitrogen. Because of  the complex structure of the N$_2$ spectrum in the near-infrared, the theoretical modeling of this spectrum is beyond the scope of this paper.

While the total number of detected lines increases with higher MW power, the number of lines that show no noticeable problems during the analysis decreases. This is mainly due to the many blends contributing to around 90\% of the flagged lines (often in combination with other flags). Total percentage of blended lines in the line list amounts to 35\%, 47\%, and 55\% for 25\,W, 50\,W, and 100\,W, respectively.

\begin{table*}
\caption{Explanations for the line list columns.}
\label{tab: listexplanation}
\centering
\begin{tabular}{rlll}
\hline \hline
Column &  Symbol\tablefootmark{a} & Unit & Explanation \\
\hline
1       &                               &               & designation of atom or molecule; ``NoID'' if not identified \\
2       & $\tilde{\nu}_{0\text{, c}}$   & cm$^{-1}$     & line center, calibrated as described in Sect.~\ref{sec: kfactor}\\
3       & $\tilde{\nu}_0$               & cm$^{-1}$     & line center\\
4       & $\sigma(\tilde{\nu}_0)$       & cm$^{-1}$     & error on $\tilde{\nu}_0$, derived from weighted line fitting (``0.'' if fit parameter is at boundary condition)\\
5       & $I_\text{c}$                  & a.u.          & line intensity, with flat-field correction as described in Sect.~\ref{sec: corrff} \\
6       & $I$                           & a.u.          & line intensity\\
7       & $\sigma(I)$                   & a.u.          & error on $I$, derived from weighted line fitting (``0.'' if fit parameter is at boundary condition)\\
8       & $\sigma_\text{G}$             & cm$^{-1}$     & width of the Gaussian line profile \\
9       & $\sigma(\sigma_\text{G})$     & cm$^{-1}$     & error on $\sigma_\text{G}$, derived from weighted line fitting  (``0.'' if fit parameter is at boundary condition)\\
10      & FWHM$_\text{c}$               & cm$^{-1}$     & full width at half maximum, corrected for instrumental broadening as described in Sect.~\ref{sec: corrlinewidth}\\
11      & FWHM                          & cm$^{-1}$     & full width at half maximum\\
12      & $\sigma(\text{FWHM})$         & cm$^{-1}$     & error on FWHM, derived by error propagation from $\sigma(\sigma_\text{G})$ \\
13      &                               & cm$^{-1}$     & distance of $\tilde{\nu}_{0\text{, c}}$ to reference line in literature (``99.'' for not identified lines) \\
14      & flags                         &               & ``x'': fit has failed or a fit parameter is at boundary condition. \\
        &                               &               & ``y'': fit residuals exceed at least 11 times (= intensity threshold+noise value) the noise \\
        &                               &               & level and are larger than 5\% of the peak intensity. \\
        &                               &               & ``b'': the neighboring line is closer than the sum of their FWHM and has more than \\
        &                               &               & half of the line's peak intensity $I$ (blend). \\
        &                               &               & ``r'': FWHM is narrower than the resolution set for the FTS. \\
        &                               &               & ``f'': $\sigma(\text{FWHM})$ is more than five time larger than the average error on FWHM (for\\
        &                               &               &   unidentified lines) or FWHM$_\text{c}$ scatters more than five times from a linear trend for \\
        &                               &               & its species (for identified lines). \\
15      &                               &               & transition as identified from \citet{Effantin1979} or \citet{Ferguson1992}.\\
\hline
\end{tabular}
\tablefoot{
\tablefoottext{a}{as used in equations and text in this article.}
}
\end{table*}

\begin{table}
\caption{Number of spectral lines in the three spectra.}
\label{tab: results}
\centering
\begin{tabular}{l|rrr}
\hline \hline
                                                & 25\,W         & 50\,W         & 100\,W  \\
\hline
Total number of lines                           & 40408         & 51776         & 58274           \\
\quad thereof lines without flags               & 24585         & 24863         & 22723           \\
\hline
identified N I lines\tablefootmark{a}           & 25            & 27            & 29              \\
\quad thereof lines without flags               & 17            & 14            & 17              \\
identified N$_2$ lines\tablefootmark{b}         & 860           & 953           & 1003            \\
\quad thereof lines without flags               & 632           & 636           & 611             \\
identified N$_2^+$ lines\tablefootmark{c}       & 886           & 898           & 903             \\
\quad thereof lines without flags               & 542           & 476           & 395             \\
unidentified lines                              & 38637         & 49898         & 56339           \\
\quad thereof lines without flags               & 23394         & 23737         & 21700           \\
\hline
\end{tabular}
\tablefoot{
\tablefoottext{a}{Out of 29 lines from the NIST database.}
\tablefoottext{b}{Out of 1260 lines of the \mbox{$(0-0)$} band of the B$^3\Pi_\text{g}$-A$^3\Sigma_\text{u}^+$ system of $^{14}$N$_2$ from \citet{Effantin1979}.}
\tablefoottext{c}{Out of 1017 lines of the \mbox{$(0-0)$}, \mbox{$(0-1)$}, \mbox{$(1-0),$} and \mbox{$(1-2)$} bands of the A$^2\Pi_\text{u}$-$X ^2\Sigma_\text{g}^+$ system of $^{14}$N$_2^+$ from \citet{Ferguson1992}.} See main text and Table~\ref{tab: listexplanation} for the meaning of flags.
}
\end{table}

\subsection{Line intensities}

\begin{figure}
\resizebox{\hsize}{!}{\includegraphics{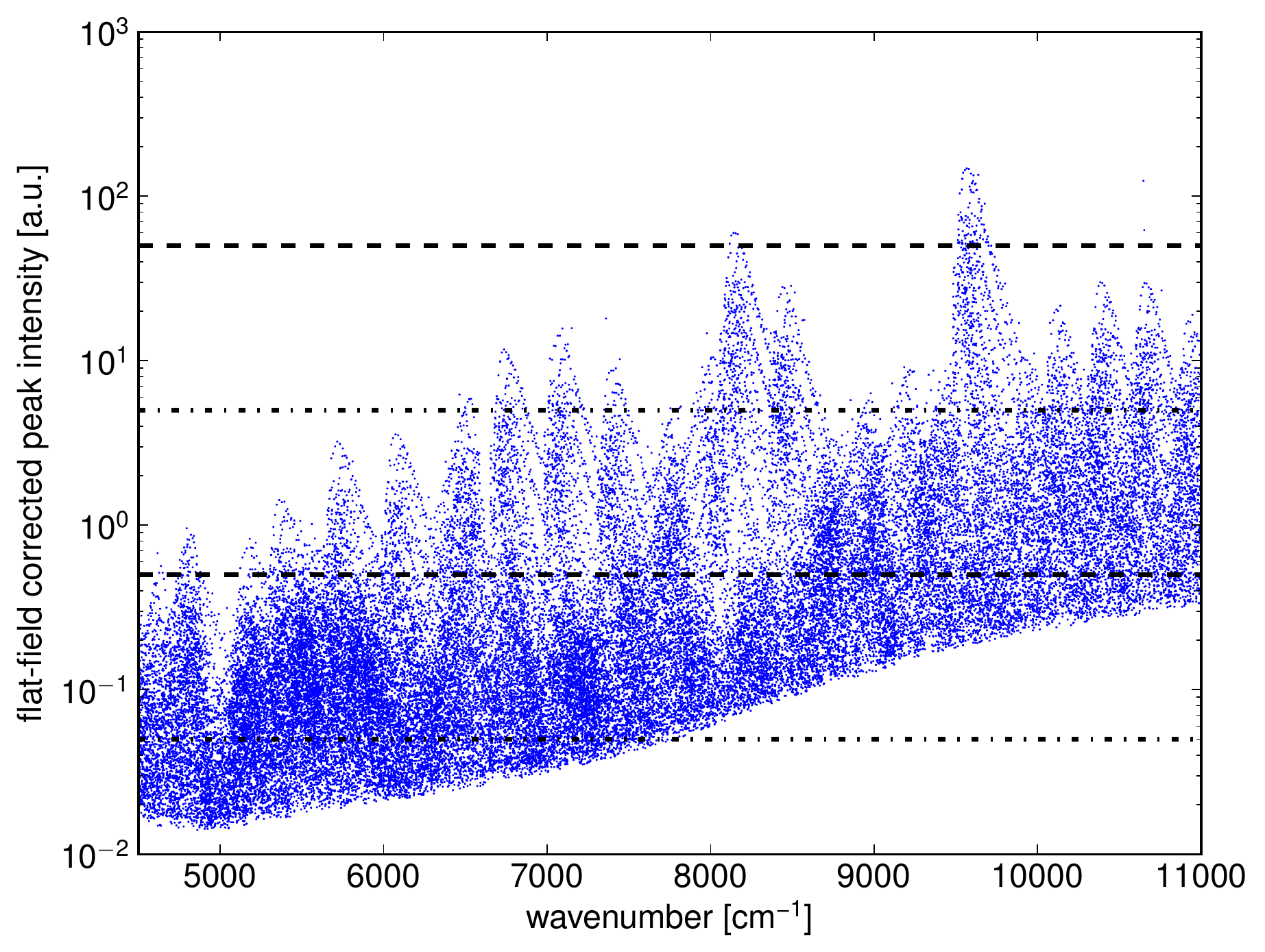}}
\caption{Line intensities after flat-field correction as a function of wavenumber. Note the logarithmic scale. The horizontal lines indicate intensity ranges used in the upper panel (dashed lines) and lower panel (dash-dotted lines) of Fig.~\ref{fig: histR}. Data from discharge operated at MW power of 50\,W.}
\label{fig: intens}
\end{figure}

We present the distribution of line intensities using the spectrum of the discharge operated at MW power of 50\,W.
Figure~\ref{fig: intens} displays the flat-field corrected line intensities $I_\text{c}$ as a function of wavenumber. The corresponding raw spectrum is shown in Fig.~\ref{fig: spectrum}. At lower intensities, the distribution exhibits an envelope with a shape related to the applied flat-field curve (compare with Fig.~\ref{fig: flat-field}). This means that our measurements are less sensitive to fainter lines toward larger wavenumbers, owing to the response curve of the InSb detector. It can therefore be expected that all emission lines with a peak intensity $I_\text{c} \geq 0.4$\,a.u. are included in our spectra and line lists, but that the sample is incomplete for fainter lines.

\begin{figure}
\resizebox{\hsize}{!}{\includegraphics{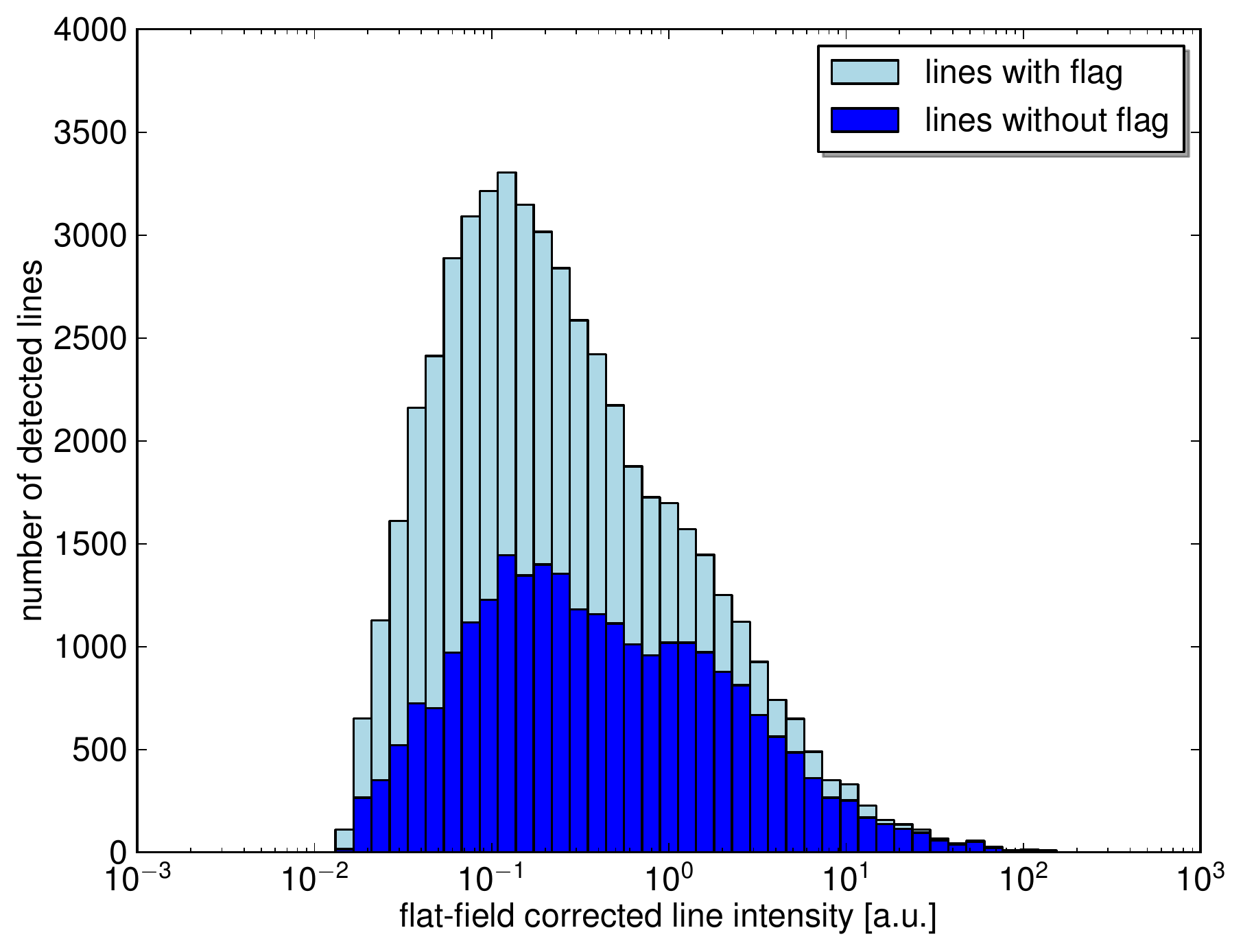}}
\caption{Histogram of the peak intensities $I_\text{c}$ of the detected lines. Bars are stacked on top of each other. Data from discharge operated at MW power of 50\,W.}
\label{fig: histintens}
\end{figure}

A histogram of the peak intensities $I_\text{c}$ is shown in Fig.~\ref{fig: histintens}, again exemplary for the discharge operated at MW power of 50\,W. The distribution peaks at around 0.1\,a.u. and decreases quickly toward smaller intensities\footnote{The designation for arbitrary units, a.u., for the intensity is dropped in the remainder of the article for better readability.}. The shape of the distribution at intensities below 0.4 is biased by the wavenumber-dependent detection efficiency. The total dynamic range of the spectral lines covers about four orders of magnitude. Figure~\ref{fig: histintens} also highlights the fraction of lines with and without flags relative to the total number of detected lines as a function of intensity. For all three measurements, the percentage of flagged lines stays above 40\% for the lower peak intensities between 0.01 and 0.1, and decreases to a value below 25\% for intensity values above 10.

Near-infrared detectors used for astrophysical spectrometers can usually distinguish line intensities over a dynamic range of two orders of magnitude. Therefore, only a fraction of the spectral lines of the molecular nitrogen spectrum are usable for wavelength calibration of astrophysical spectrographs in practice. However, as the utilization of a spectral line of molecular nitrogen for wavelength calibration should not depend on its intensity, an observer is flexible in selecting the appropriate exposure time for a certain calibration frame. 

\subsection{Line density}

\begin{figure}
\resizebox{\hsize}{!}{\includegraphics{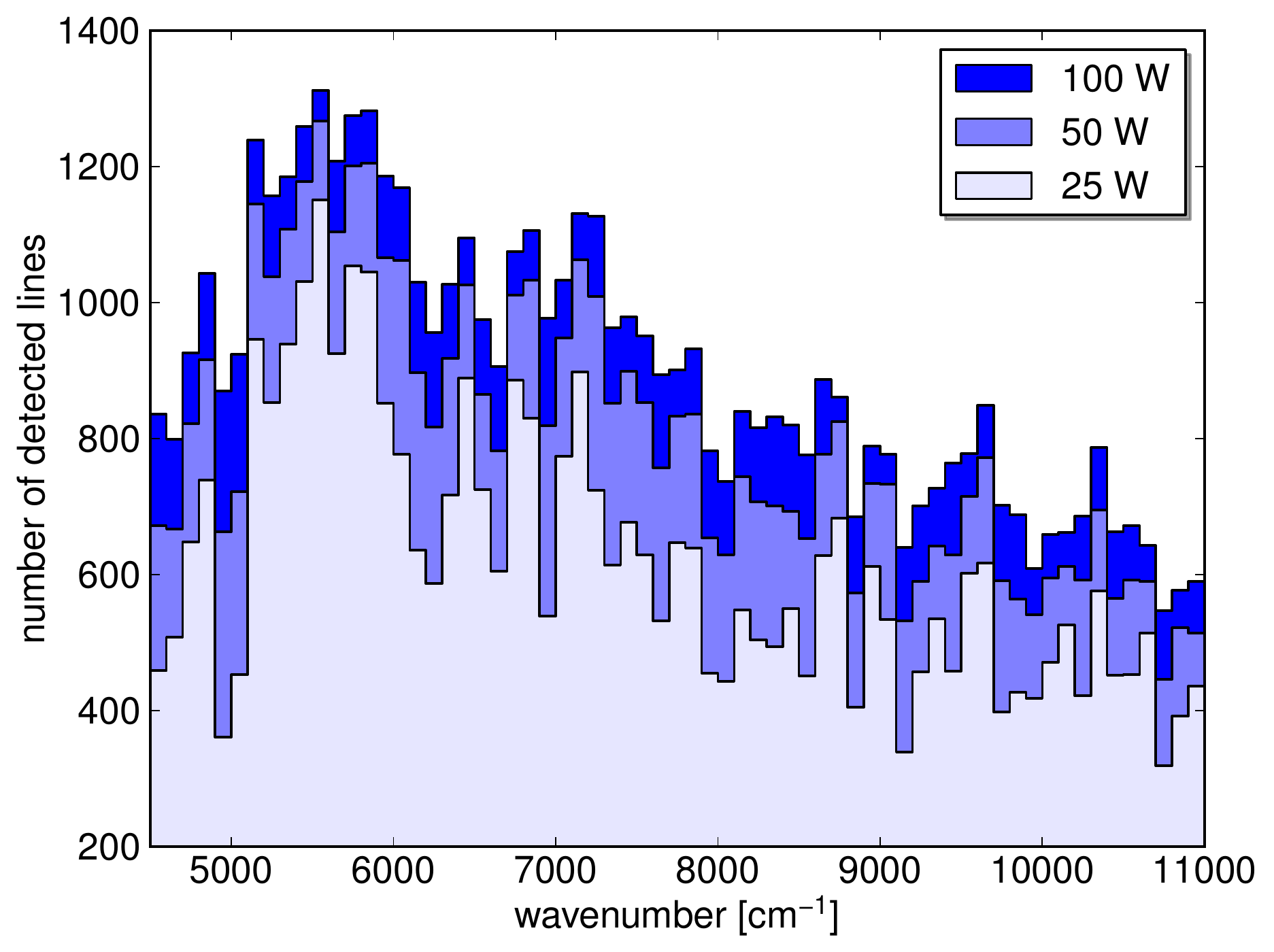}}
\caption{Histograms showing the distribution of emission lines as function of wavenumber. Each bin covers a range of 100\,cm$^{-1}$. The bars are not stacked and show the absolute number of lines in one bin for each of the three spectra operated at different MW powers.}
\label{fig: wnhist}
\end{figure}

The spectrum of the nitrogen gas discharge exhibits emission lines over the complete spectral range under investigation. A histogram of the line distribution is shown in Fig.~\ref{fig: wnhist} for all three measurements. Around 350 to 1300 lines are located within one bin covering 100\,cm$^{-1}$. The number of lines in each bin increases with higher operating MW power.

\begin{figure}
\resizebox{\hsize}{!}{\includegraphics{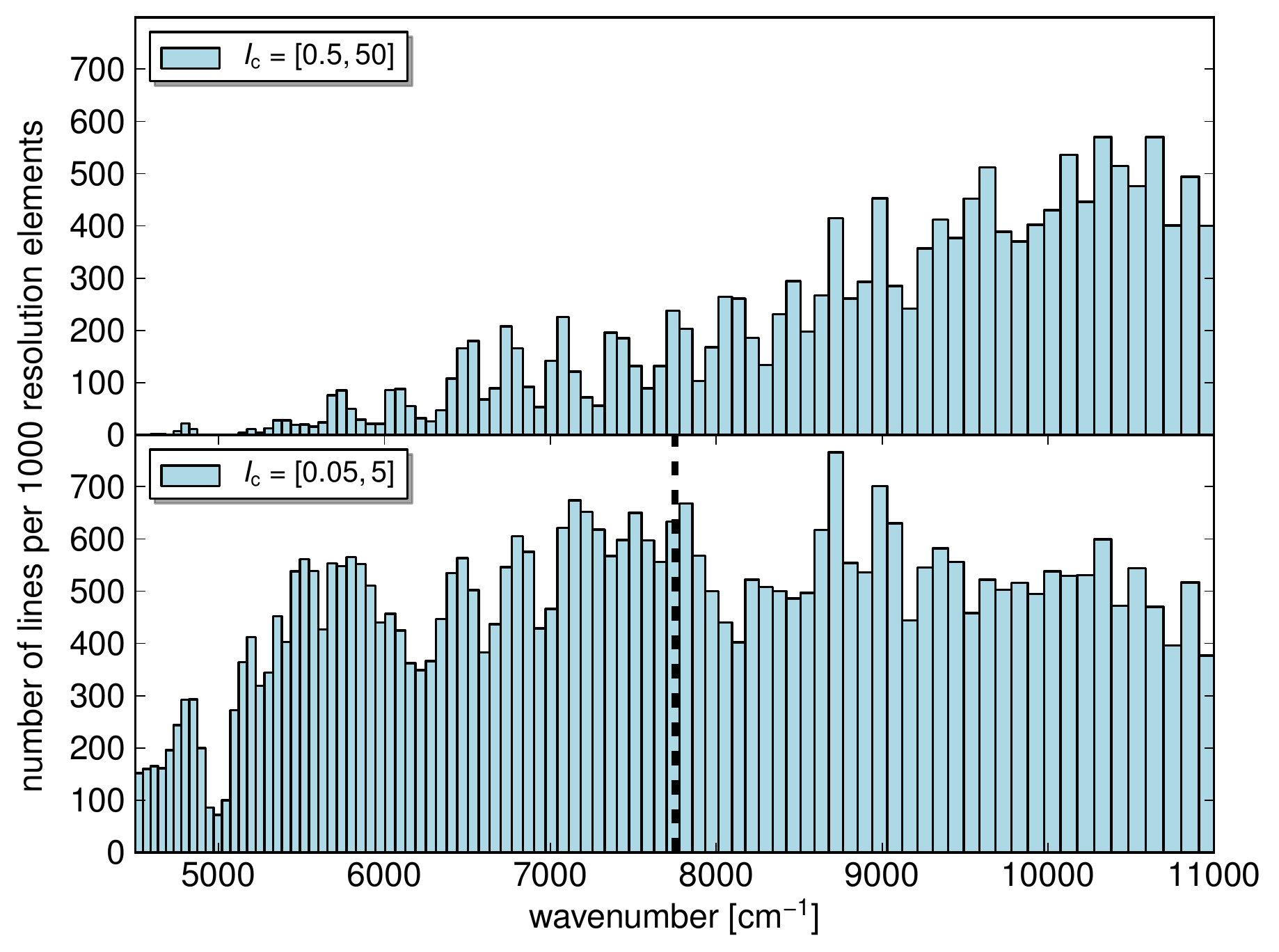}}
\caption{Histograms of detected emission lines as a function of wavenumber. The samples are limited to emission lines with a peak intensity $0.5 - 50$ (upper panel) and $0.05 - 5$ (lower panel). Each bin covers 1000 resolution elements based on a resolution of $R = 100\,000$. In the lower panel, the dashed vertical line is located at 7750\,cm$^{-1}$ and the sample is expected to be incomplete at higher wavenumbers (compare with Fig.~\ref{fig: intens}). Data from discharge operated at MW power of 50\,W.}
\label{fig: histR}
\end{figure}

In order to be useful for wavelength calibration, a spectrum needs a sufficient number of lines, depending on the instrument's resolution and wavelength coverage. Similar to Fig.~\ref{fig: wnhist}, Fig.~\ref{fig: histR} shows the distribution of lines as a function of wavenumber, but this time each bin covers a constant number of 1000 resolution elements with a resolution of $R=\lambda/\Delta \lambda = 100\,000$. This corresponds to specifications of high-resolution grating spectrographs like, e.g., CRIRES \citep{Kaeufl2004}. Considering the expected dynamic range of a near-infrared detector, in Fig.~\ref{fig: histR} we only include emission lines with peak intensities within two orders of magnitude: the histogram in the upper panel of Fig.~\ref{fig: histR} contains lines with flat-field corrected intensities of $0.5-50$ (compare horizontal lines in Fig.~\ref{fig: intens}). For example, at wavenumbers above 9000\,cm$^{-1}$, an average of about 0.4 lines fall into one resolution element. This number decreases toward smaller wavenumbers until very few lines of this intensity are available at 4500\,cm$^{-1}$. To detect more lines at smaller wavenumbers, we need to go to lower intensities as shown in the lower panel, which can be achieved using longer exposure times and a longpass filter. The second histogram contains lines with intensities $0.05-5$. Here, on average, there are approximately 0.5 lines per resolution element in the wavenumber range $5500-7750\,$cm$^{-1}$. At wavenumbers above 7750\,cm$^{-1}$ (dashed line) fainter lines are missing in our measurements.

The continuous spread of more than 40\,000 detected lines over the wavenumber range $4500-11000\,$cm$^{-1}$ ensures a sufficient number of lines for wavelength calibration. The exposure time might need to be adjusted when observing in different spectral regions to obtain the optimal number of lines, depending on a spectrograph's specific design.

\subsection{The influence of different MW powers on line width and intensity}
\label{sec: MWPower}
Varying the microwave power supplied to the cavity changes the mean power absorbed by electrons from the electric field and therefore the rate of collisional excitations in the plasma \citep[chapt. 1.2]{Jankowski2011}. Here we present how the widths of the spectral lines become broader with higher MW power and how the intensities of the atomic nitrogen lines increase with MW power.

Doppler broadening is expected to be the dominating effect on the line width after correction for instrumental line broading because of the low gas pressure and the Gaussian-shaped spectral lines. The theoretical description of Doppler broadening can be expressed as 
\begin{align}
 \text{FWHM}_\text{c} = \sqrt{\frac{8 kT \ln(2)}{mc^2}} \tilde{\nu}_0 , \label{eq: temperature}
\end{align}
where $k$ is the Boltzmann constant, $m$ is the mass of the species, and $c$ is the speed of light \citep[e.g.,][chap. 1.3]{Bernath2005}. According to Eq.~\ref{eq: temperature}, lighter species create broader spectral lines compared to heavier species (at a constant temperature $T$), and the line width increases linearly with wavenumber $\tilde{\nu}_0$.

\begin{figure}
\resizebox{\hsize}{!}{\includegraphics{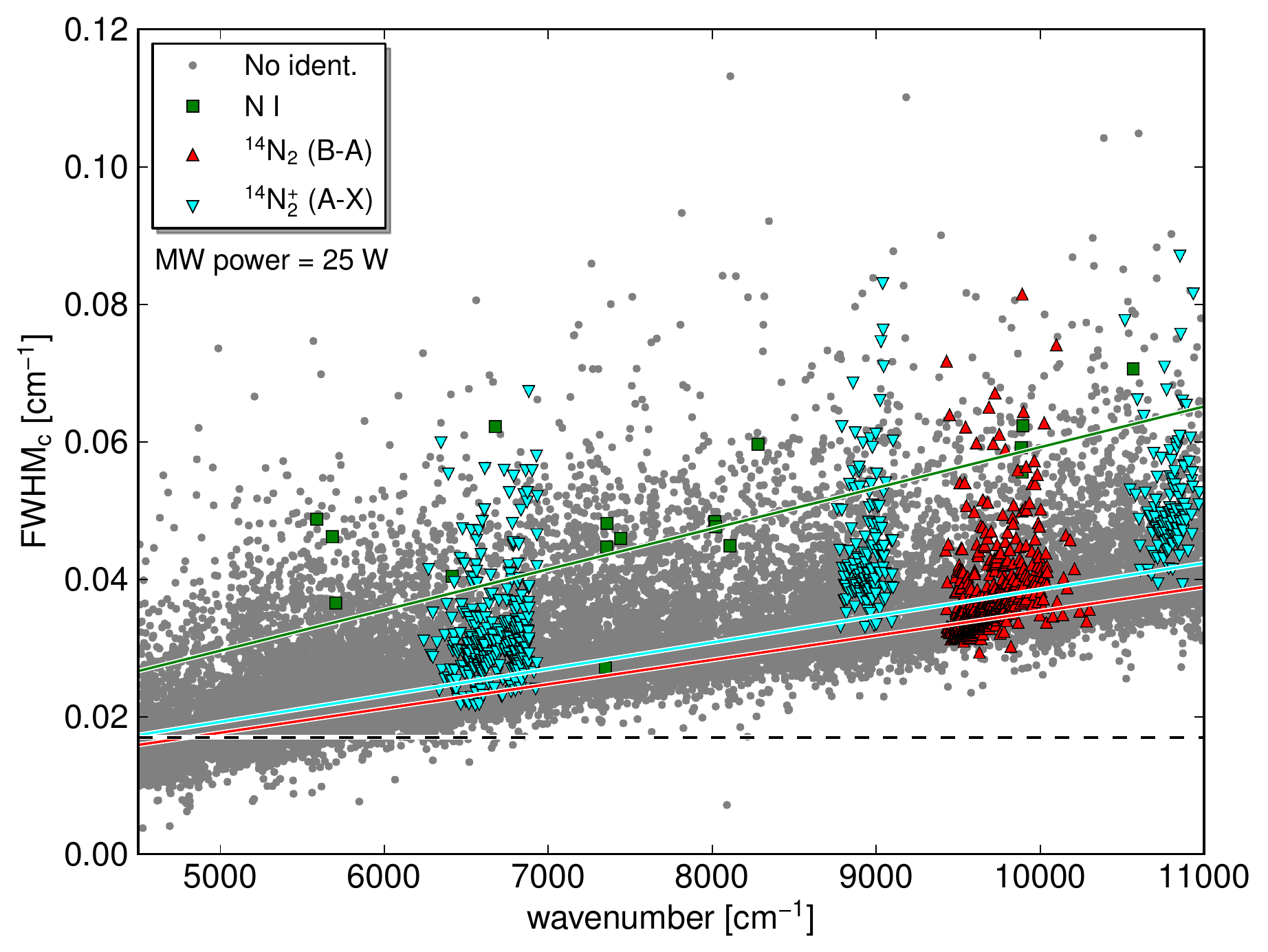}}
\resizebox{\hsize}{!}{\includegraphics{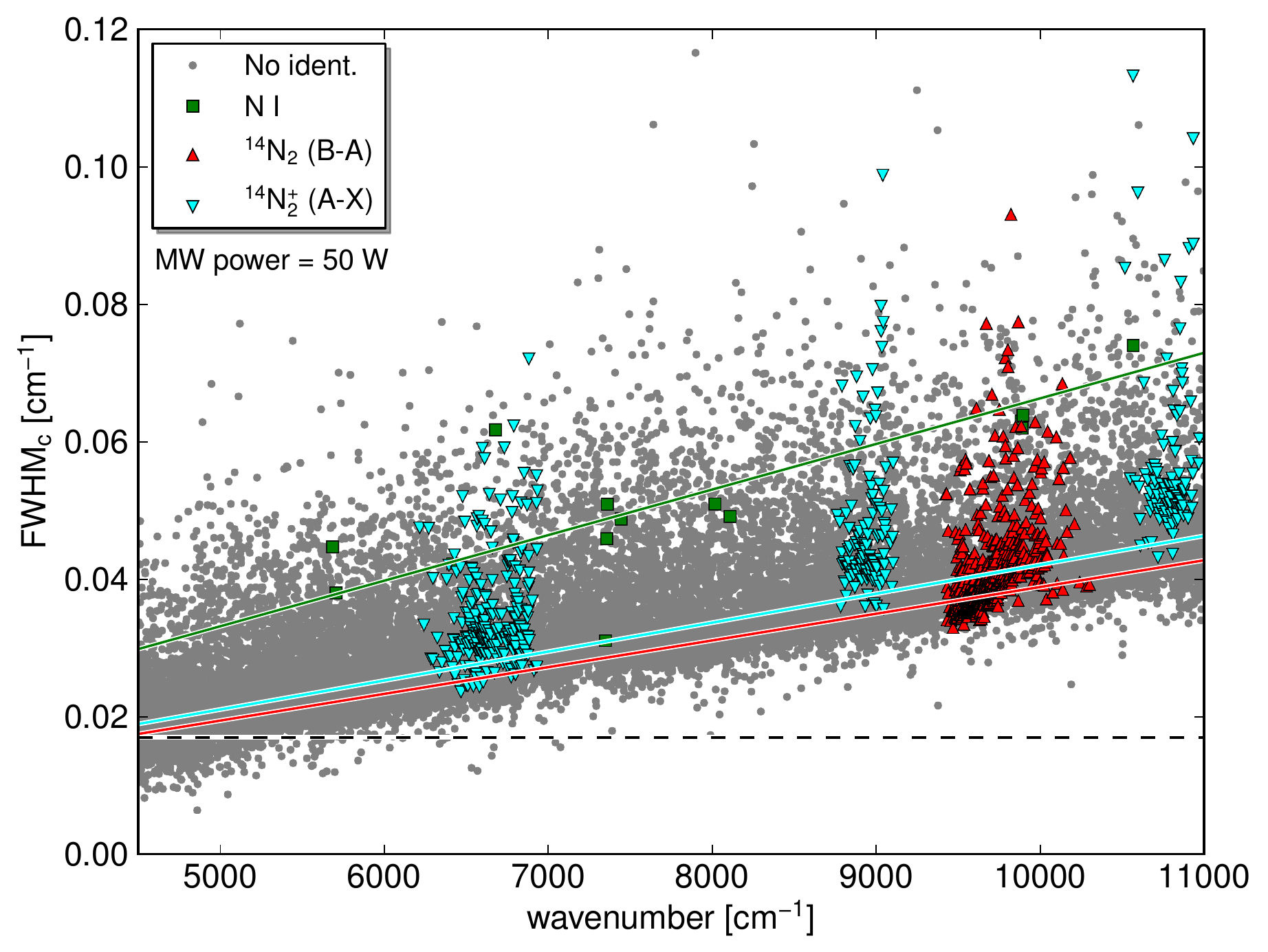}}
\resizebox{\hsize}{!}{\includegraphics{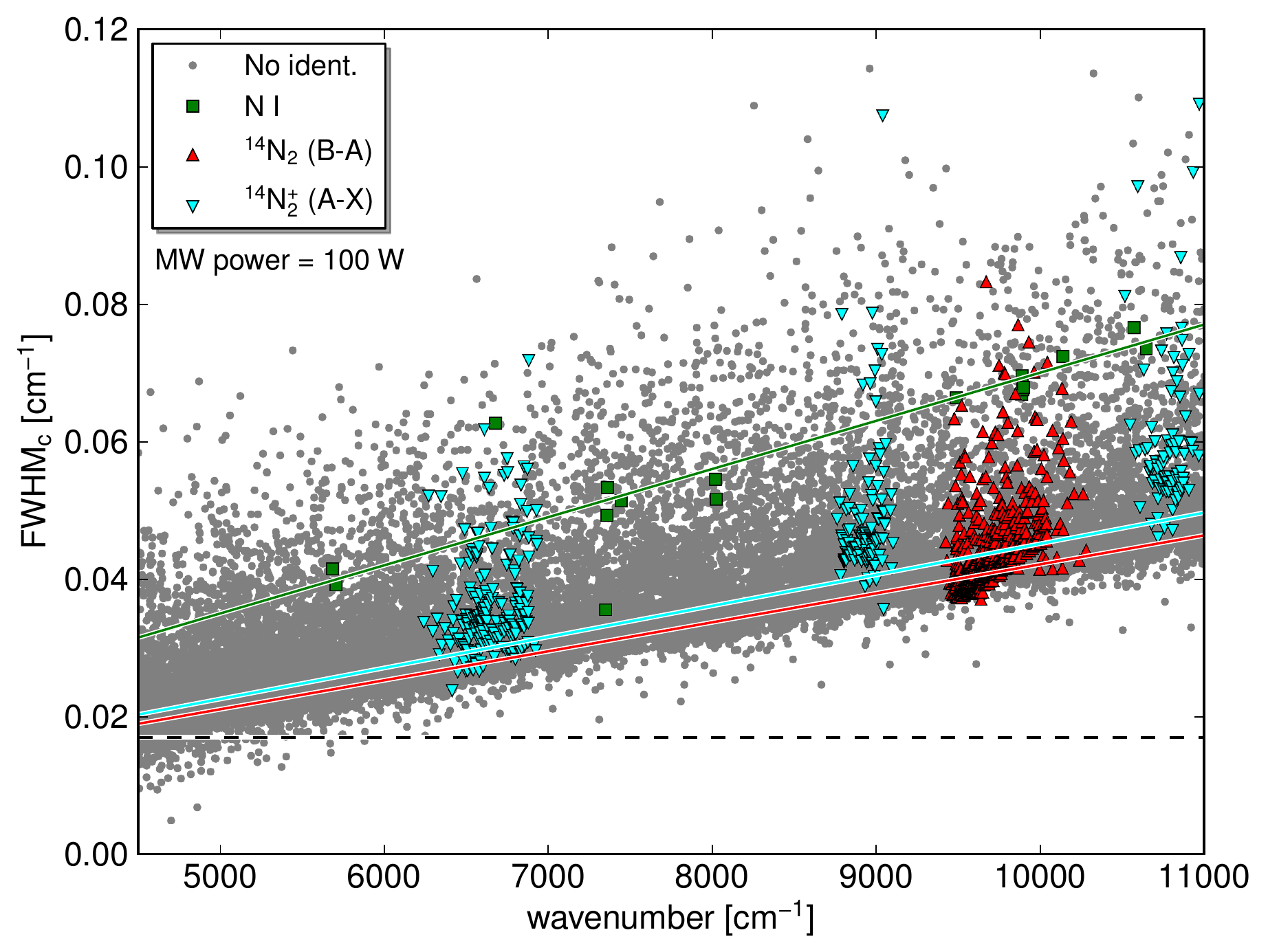}}
\caption{Each point represents the FWHM$_\text{c}$ of one spectral line as a function of wavenumber. The solid lines show linear regressions with ordinate crossing forced at the origin. The black dashed horizontal line indicates the FTS resolution. From top to bottom: discharge operated at microwave power of 25, 50, and 100\,W.}
\label{fig: FWHM}
\end{figure}

The FWHM$_\text{c}$ of all lines without a flag is plotted against wavenumber in Fig.~\ref{fig: FWHM}. The identified lines of atomic and molecular nitrogen are highlighted. A linear regression using Eq.~\ref{eq: temperature} is applied to the data points of each identified species, as represented by the lines in the respective colors in Fig.~\ref{fig: FWHM}. The linear regression weights each point according to its uncertainty in FWHM$_\text{c}$, which can lead to a slight displacement of the linear regression relative to the bulk of the data. Atomic nitrogen clearly follows a trend of larger FWHM$_\text{c}$ as compared to molecular nitrogen. We also see the expected linear trend with wavenumber for all species. The distribution of the FWHM$_\text{c}$ from the unidentified lines is consistent with a combined FWHM$_\text{c}$ distribution of lines from N$_2$ and N$_2^+$ with a ratio of $0.82:0.18$, respectively. The overall scatter is, however, too large to make conclusive statements about the origins of individual lines.

Equation~\ref{eq: temperature} includes a temperature parameter $T$. The definition of a single gas temperature is here not possible because a microwave induced plasma is not in a local thermodynamic equilibrium. Each species in the plasma rather has its own temperature \citep[chapter 1.4]{Jankowski2011}. Because a change in temperature for a certain species is related to a change in FWHM, we compile the values of $T$ as derived from the linear regression for completeness in Table~\ref{tab: temperature}. The temperature rises for all species if the MW power is increased. The error on $T,$ as obtained directly from the fit, is very small ($\leq 1$\,K) because the linear regression consists of only one free parameter.  To estimate more realistic uncertainties, we varied the data points by $\pm \sigma(\text{FWHM}_\text{c})$ and report the resulting change in $T$ in Table~\ref{tab: temperature}. 

\begin{table}
\caption{Values of the temperature parameter $T$ in Eq.~\ref{eq: temperature} as derived by linear regression.}
\label{tab: temperature}
\centering
\begin{tabular}{l|rrr}
\hline \hline
                        & N I           & N$_2$\tablefootmark{a}        & N$_2^+$\tablefootmark{b}\\
\hline
$T$(25\,W) [K]          & 957 $\pm$ 10  & 681 $\pm$ 11  & 809 $\pm$ 36 \\
$T$(50\,W) [K]          & 1200 $\pm$ 2  & 825 $\pm$ 8   & 967 $\pm$ 27 \\
$T$(100\,W) [K]         & 1339 $\pm$ 2  & 970 $\pm$ 7   & 1113 $\pm$ 20 \\
\hline
\end{tabular}
\tablefoot{
\tablefoottext{a}{1 band of the B$^3\Pi_\text{g}$-A$^3\Sigma_\text{u}^+$ system.}
\tablefoottext{b}{4 bands of the A$^2\Pi_\text{u}$-$X ^2\Sigma_\text{g}^+$ system.}
}
\end{table}

\begin{figure}
\resizebox{\hsize}{!}{\includegraphics{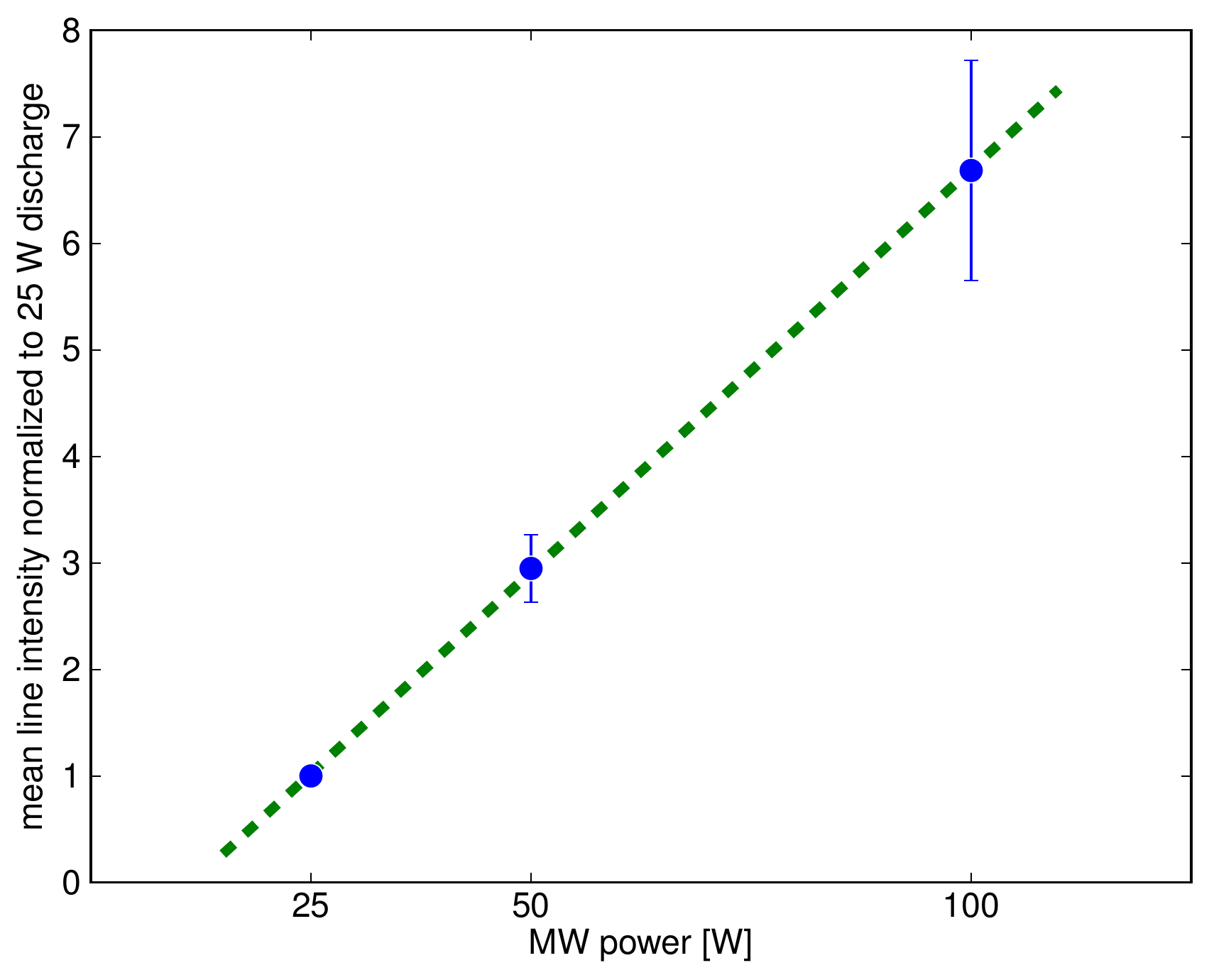}}
\caption{The mean intensity of the detected N I lines relative to the respective line intensity in the spectrum of the discharge operated at 25\,W. The green dashed line indicates a linear regression.}
\label{fig: intenschange_NI}
\end{figure}

Furthermore, we analyzed the change in peak intensities for the 25 transitions of N~I that are identified in all three spectra. Figure~\ref{fig: intenschange_NI} displays the mean change in peak intensity when the MW power is increased from 25\,W to 50\,W and then to 100\,W. The values are calculated in the following way: the intensity $I_{\text{c, }P}$ of each N~I line in the spectrum of the discharge operated at MW power $P$ is divided by the intensity $I_\text{c, 25\,W}$ of the same line in the spectrum of the discharge operated at 25\,W; then, the mean value is calculated. The latter appears to follow a linear trend as indicated by the dashed line in Fig.~\ref{fig: intenschange_NI} and described by the following linear regression:

\begin{align}
\left\langle \frac{I_{\text{c, }P}}{I_\text{c, 25\,W}} \right\rangle =  0.0756\,\frac{1}{\text{W}} \cdot P[\text{W}] - 0.8679 .
\end{align}

The individual intensity values are well separated into two disjointed distributions for the measurements with 50\,W and 100\,W, with only one outlier at 7349\,cm$^{-1}$. If we remove this outlier, the error bars shrink by about a factor of two and the linear regression becomes $\left\langle \frac{I_{\text{c, }P}}{I_\text{c, 25\,W}} \right\rangle =  0.0781\,\frac{1}{\text{W}} \cdot P[\text{W}] - 0.9318$.

\section{Conclusion}

We have presented the characterization of three emission spectra of a nitrogen gas discharge in the range $4500-11000$\,cm$^{-1}$ created by a microwave-induced plasma. A specific characteristic of the experimental setup is the sealed gas cell, which allows one to repeat the measurement with the same gas properties. The spectral measurements were carried out for three different MW powers (25\,W, 50\,W, and 100\,W). We detected and fitted 40408, 51776, and 58274 emission lines in these spectra, respectively. Spectral line lists with line parameters are provided for all detected lines. The spectra exhibit densely-spaced lines throughout the whole wavenumber range under investigation with about $350-1300$ lines per 100\,cm$^{-1}$. Depending on the MW power, between 35\% and 55\% of all lines are blended. We identified molecular and atomic nitrogen lines using existing line lists from the literature. The majority of the lines remains unidentified. Theoretical transition wavenumbers are needed to confirm whether they originate from nitrogen and to properly assign all transitions.

The analysis of line intensities and line density focused on the potential application as a wavelength calibration source for astrophysical spectrographs. The number and distribution of emission lines qualifies the spectrum of the recorded source to be used for wavelength calibration over the whole spectral range under investigation. For a resolution of $R=100\,000$, approximately one line every two resolution elements can be realized over a spectral range of a few thousand wavenumbers with peak intensities within two orders of magnitude. Depending on the wavelength range of a specific spectrometer, exposure times need to be adjusted to select, in combination with a bandpass filter, a suitable subset of emission lines. Tests for stability and aging effects need to be carried out to further investigate the feasibility of the molecular emission spectrum as a reliable wavelength calibrator.

\begin{acknowledgements}
We thank Philipp Huke for assisting with the measurements of the frequency-locked laser and Ulf Seemann for helpful comments on the manuscript. We acknowledge financial support by the European Research Council under the FP7 Starting Grant agreement number 279347 and by the DFG Research Training Group 1351 ``Extrasolar Planets and their Host Stars''. AR acknowledges funding through a Heisenberg professorship under DFG RE 1664/9-2.
\end{acknowledgements}


\bibliographystyle{aa} 
\bibliography{N2spectrum_bibliography}

\end{document}